\documentclass[preprint,12pt,sort&compress]{elsarticle}
\graphicspath{{./Fig/}}
\usepackage{amssymb}
\usepackage{amsthm}
\usepackage{amsfonts}
\usepackage{amsmath,bm} 
\usepackage{gensymb}	
\usepackage{graphicx}
\usepackage{xcolor}			   
\usepackage{dcolumn}
\usepackage{booktabs}

\usepackage{lmodern}           
\usepackage[T1]{fontenc}       
\usepackage[cp1250]{inputenc}  
\usepackage{verbatim} 	       
\usepackage{setspace}
\journal{Composite Structures}

\usepackage[demo]{adjustbox}
\usepackage{xcolor}
\usepackage{atbegshi}
\AtBeginDocument{\AtBeginShipoutNext{\AtBeginShipoutDiscard}}	
\PassOptionsToPackage{hyphens}{url}\usepackage{hyperref} 

\begin{document}
	\pagestyle{empty} 
	\begin{titlepage}
		\color[rgb]{.4,.4,1}
		\hspace{5mm}

		\bigskip
		
		\hspace{15mm}
		\begin{minipage}{10mm}
			\color[rgb]{.7,.7,1}
			\rule{1pt}{226mm}
		\end{minipage}
		\begin{minipage}{133mm}
			\vspace{10mm}        
			\color{black}
			\sffamily
			\LARGE\bfseries Nonlocal integral thermoelasticity:  \\[-0.3\baselineskip] a thermodynamic framework  \\[-0.3\baselineskip] for functionally graded beams
			
			\vspace{5mm}
			{\large {Preprint of the article published in \\[-0.4\baselineskip] Composite Structures \\[-0.1\baselineskip] 225, 1 October 2019, 111104 }} 
			
			\vspace{10mm}        
			{\large Raffaele Barretta,\\[-0.4\baselineskip] \textsc{Marko \v{C}ana\dj{}ija}, \\[-0.4\baselineskip] Francesco Marotti de Sciarra} 

			\large
			
			\vspace{40mm}
   			\vspace{5mm}
   			
			\small
			\url{https://doi.org/10.1016/j.compstruct.2019.111104}
			
			\textcircled{c} 2019. This manuscript version is made available under the CC-BY-NC-ND 4.0 license \url{http://creativecommons.org/licenses/by-nc-nd/4.0/}
			\hspace{30mm} 
			\color[rgb]{.4,.4,1} 
		\end{minipage}
	\end{titlepage}

\begin{frontmatter}

\title{Nonlocal integral thermoelasticity: a thermodynamic framework for functionally graded beams}

\author[UN]{Raffaele Barretta}
\author[URI]{Marko \v{C}ana\dj{}ija \corref{cor}}
\ead{marko.canadija@riteh.hr}
\cortext[cor]{Corresponding author. Tel.: +385-51-651-496; Fax.: +385-51-651-490}
\author[UN]{Francesco Marotti de Sciarra}

\address[UN]{Department of Structures for Engineering and Architecture, University of Naples Federico II, Via Claudio 21, 80121 Naples, Italy}
\address[URI]{University of Rijeka, Faculty of Engineering,Department of Engineering Mechanics, Vukovarska 58, 51000 Rijeka, Croatia}

\begin{abstract}
An effective nonlocal integral formulation for functionally graded Bernoulli-Euler beams in nonisothermal environment is developed. Both thermal and mechanical loadings are accounted for. The proposed model, of stress-driven integral type, is shown to be governed by a thermodynamically consistent differential problem with proper constitutive boundary conditions. The new thermoelastic strategy is illustrated by investigating a set of examples. It is demonstrated that in nonisothermal statically indeterminate problems rather complex structural behaviours can appear and that both the shift of the neutral surface and nonlocality have a dominating influence at small-scales.
\end{abstract}

\end{frontmatter}

\section{Introduction}
Beam models that can capture small size behaviour, typically observed in very small devices, is a very active research field nowadays. As small beams, we presume structures having dimensions of micrometer or nanometer order. Size effects (i.e. nonlocal phenomena) are observed since the forces irrelevant at the classical engineering scale (where structures are sized in centimetres or meters) are now important in such small structures. Increased technological developments at microscale and nanoscale require more accurate mechanical solutions, fuelling the increased interest of the research community \cite{Chen2019,Chen2018,Zhu2019}.

Initial contributions to the solution of these problems were based on the gradient formulation. Although some paradoxical results were reported rather early in literature \cite{Challamel08,Peddieson2003}, such an approach remains a popular choice. The paradox has puzzled scientists for quite some time, and some attempts based on the strain-driven integral formulations were proposed in literature. Unfortunately, some strain-driven models suffer from the problem of fulfilling equilibrium conditions \cite{Romano17}. On the contrary, integral formulations of the stress-driven type \cite{Romano2017c} are able to properly address issues relating to the fulfilment of equilbrium conditions and paradoxes debated in the scientific community \cite{Romano17b,Barretta2018b}. Moreover, it is pointed out that, at least for a certain class of kernel function, the stress-driven integral formulation can be made equivalent to differential problems \cite{Barretta2018d,Barretta2018c,Barretta2019}. To ensure this equivalence, suitable constitutive boundary conditions must be set in the correct manner. These are different in isothermal and nonisothermal problems.

Nonisothermal environmental conditions can have a profound influence on the mechanical response of nonlocal beams. In the vibrational context, temperature has a significant influence on the frequency shift \cite{Ghaffari2018}. Thus, a good part of research efforts on nonlocal beams in nonisothermal environments are devoted to dynamical analysis \cite{Malikan2018}. Composite beams in this context are analysed in \cite{Barati2018,Mahmoudpour2018,Saffari2017,Shafiei2018,Rahmani2017}. These analyses tend to account for other phenomena and become rather complex, like in the thermo-electro-mechanical analysis of functionally graded size dependent beams \cite{Jia2018}. Non-probabilistic uncertainty modelling for vibration and buckling of the FG nanobeams in nonisothermal conditions is considered in \cite{Lv2018}. In the statical analysis, buckling of nonlocal functionally graded beams under thermal loading is frequently addressed, starting from the early contribution in \cite{Murmu09} till more recent contributions \cite{She2017,Ebrahimi2017,DehrouyehSemnani2017,Semmah2015,Elmerabet2017}. Mainly nonlocal influence on the mechanical part of the problem is investigated, but buckling caused by the size effect on heat conduction is also analysed \cite{Yu2016}. Another recent contribution \cite{Dehrouyeh2017} points out the difference in  original and simplified boundary conditions in vibrational and buckling analyses of FG beams in nonisothermal environments. Original boundary conditions do include the thermal moment at the flexural boundary conditions. The obtained results clearly demonstrate significant differences of two types of boundary conditions. Bending of thermoelastic nonlocal beams were analysed in \cite{Canadija16a, Canadija16b}.  Majority of these results are based on gradient formulations, both for Bernoulli-Euler and Timoshenko beams. Only a few solutions based on the integral formulation are available, see \cite{Canadija2018} for an example.

It is noteworthy that in composite beams the neutral surface does not coincide with the geometrically neutral surface \cite{She2018, She2019}. This is a direct consequence of the functional grading of beam's material. Such effects does bring additional level of complexity to the modelling procedure. Even in the classical, local case the most frequent approach is to disregard these effects. Nevertheless, some results including this shift in local problems do exists. For the present research, nonisothermal model \cite{Dehrouyeh2017} is of interest. But, when the nonlocal procedures are concerned, inclusion of the shift of neutral surface is much less common; for the isothermal strain gradient formulation see \cite{Al2015,Eltaher2013,Larbi2013,Ding2018} and for the integral based \cite{Barretta2018e}. For the gradient based thermoelastic beams including hygro effects, see \cite{Barati2018}. In the case of stress-driven nonlocal thermoelasticity, these are practically non-existent. This is surprising, since its level of influence on the results should be at least the same as the small size effects. Thus, this paper aims to include such effects into a thermoelastic formulation.

To summarize, novelty of the present paper is inclusion of the neutral surface shift into the formulation by extending previous results obtained for the nonlocal nonisothermal Bernoulli-Euler formulation \cite{Canadija2018} in the case of homogeneous beam. As a consequence, the formulation obtained in such manner is suitable for beams made of functionally graded materials. It is also shown that the strict thermodynamic framework is imperative in these derivation. Starting from the first and second law of thermodynamics, a suitable Gibbs potential is developed including all constraints arising from the stress-driven integral nature of the model. Minimization of this newly proposed potential yields the system of ordinary differential equations governing the mechanical process. The proposed formulation is thoroughly tested on four examples.

The paper is organized as follows: in order to provide the notation and a stage for further developments, the Bernoulli-Euler beam kinematics is presented at the start. The next section describes the transformation from the integral law to the equivalent differential formulation. Subsequently, to accommodate this formulation, a strict thermodynamic framework is introduced. A cornerstone of the formulation is the new potential; minimization of the Gibbs potential provides governing equations as well as the complete set of boundary conditions. The example section is followed by the conclusions that summarize main findings and close the paper.

\section{Beam kinematics in the nonisothermal setting} 
To keep elaborations as simple as possible, plane bending of an initially straight Bernoulli-Euler beam will be considered. The longitudinal axis of the beam is denoted as $x$, while the bending is assumed to take place in the $x-z$ plane. The z-axis is assumed to originate at the geometric centroid of the cross-section, so that the first moment of area vanishes $\int_{\Omega} z \mathrm{d}A=0$. The beam is made of functionally graded (FG) material, where grading does take place along $z$ axis. 

In the case of a FG beam, the physical neutral surface, in which the normal stress vanishes, does not coincide with the geometrical middle surface. The issue is recently raised in \cite{Dehrouyeh2017} for thermally loaded FG local beams. Thus, in calculation of the axial displacement, the shift of neutral surface from the geometrical middle surface $z_0$ must be accounted for. It will be demonstrated that $z_0$ can vary along the beam if $z_0=z_0(x)$. Nevertheless, the present research will not address such problems.

The central task is to determine the Cartesian components of the displacement vector field. Having in mind the remark about the shift of neutral surface, these are \cite{Dehrouyeh2018}:
\begin{equation}
\label{eq:DispField}
u_x(x,z)=u(x,z)=u_0(x)+\varphi(x) (z-z_0), \quad u_y=0  , \quad u_z(x)=w(x).\end{equation}
Above, $\varphi(x)$ denotes the angle of rotation of the cross section. The longitudinal displacement $u_x$ is composed of two parts. The first part $u_0$ represents the average displacement of the cross section, defined as the integral of the displacement over the cross sectional domain $\Omega$:
\begin{equation}\label{eq:Avgu}
u_0(x)=\frac{1}{A}\int_\Omega u(x,z) \mathrm{d}A,
\end{equation}
where $A$ denotes the cross sectional area. In the Bernoulli-Euler formulation cross sections remain orthogonal to the beam axis. This implies vanishing shear strains $\gamma _{xz}$ as the sum of the derivative of longitudinal and transversal displacement:
\begin{equation}\label{eq:PhiDispRelation}
\gamma _{xz} (x)=0=\partial _x w(x)+\partial _z u(x,z)=w^{(1)}(x) + \varphi(x),
\end{equation}
providing the link between the transversal displacement and angle of rotation:
\begin{equation}\label{eq:PhiDispRelation2}
w^{(1)}(x) = -\varphi(x).
\end{equation}
The apex $^{(n)}$ denotes the $n$-th derivative with respect to the longitudinal coordinate $x$. The axial strain in a point follows from the differentiation of the axial displacement field with respect to $x$:
\begin{equation}\label{eq:Epsilon}
\varepsilon (x,z)=\partial _x u(x,z)=u_0^{(1)} +\varphi ^{(1)}(z-z_0). \\
\end{equation}
Such a kinematic framework accounts for both isothermal and nonisothermal effects.  

It remains to explicitly introduce the temperature effects. In order to do so, the normal strain is additively separated into a thermal part $\varepsilon_{\mathrm{T}}=\alpha \Delta \theta$ and a mechanical part $\varepsilon_{\mathrm{M}}$:
\begin{equation}\label{eq:StrainAddSep}
\varepsilon=\varepsilon_{\mathrm{T}}+\varepsilon_{\mathrm{M}},
\end{equation}
where $\alpha$ is the coefficient of thermal expansion and $\Delta \theta (x,z)$ is the temperature change field. With Eqs.~(\ref{eq:Epsilon},\ref{eq:StrainAddSep}), the mechanical strain is:
\begin{equation}\label{eq:EpsM}
\varepsilon_{\mathrm{M}}=\varepsilon-\alpha \Delta \theta =u_0^{(1)} -w ^{(2)}(z-z_0)-\alpha \Delta \theta.
\end{equation}
In the local formulation, the mechanical strain is the source of normal stress, whereas in the nonlocal formulation its determination is not that much straightforward.

\section{Thermodynamic preliminaries for functionally graded beams}\label{sec:Thermodynamic}

\subsection{Motivation}
The cornerstone of the FG formulation is the stress-driven nonlocality defining constitutive law for the mechanical part of axial strain:
\begin{equation}\label{eq:MotStrain}
\varepsilon_\mathrm{M}(x,z)=\varepsilon-\alpha\Delta\theta= \int_{0}^{L} \phi_\mathrm{c} ({x-\xi)}  E^{-1} {\sigma}(\xi,z) \mathrm{d} \xi,
\end{equation}
where the kernel function $\phi_\mathrm{c} ({x)}$ is
\begin{equation}\label{eq:Kernel}
\phi_\mathrm{c}(x) = \frac{1}{2 L_\mathrm{c}} \exp(-\frac{\left| x\right| }{ L_\mathrm{c}}),
\end{equation}
${\sigma}$ is the  axial stress and $E$ is the Euler-Young's modulus. The characteristic length is defined as $L_\mathrm{c}= c L$, i.e. as the product of the small-size parameter $c$ and the beam's length $L$. Note that such assumption accounts only for nonlocality with respect to the longitudinal coordinate. Due to the exponential nature of the kernel function, the nonlocal effects in the vicinity of the corresponding point have significantly more influence on the strain than points that are situated at a larger distance. 

It should be emphasized that the precise value of the small size parameter still remains an open question. Preferably, the value should be determined experimentally, but only rare attempts to quantify size-dependent behaviour can be found in the literature, see \cite{AbuAl-Rub2004,Perkins1973,Kakunai1985} for an insight. Available numerical calculations \cite{Canadija17,Zhang09,Arash10,Narendar11,Duan07,Xiao06} also indicate that further work remains to be done in order to solve the issue.

If one wishes to obtain stresses from known strains, it can shown that the solution and necessary boundary conditions follow from the differential formulation \cite{Polyanin98,Canadija2018}:
\begin{equation}\label{eq:MotStress2}
\begin{array}{l}
\frac{{\sigma}}{E}   = -L_\mathrm{c}^{2} {(\varepsilon-\alpha\Delta\theta)}^{(2)} + \varepsilon-\alpha\Delta\theta,
\end{array}
\end{equation}
or alternatively in a more compact form:
\begin{equation}\label{eq:MotStress2DispEpsM}
\begin{array}{l}
\sigma   = E\left( -L_\mathrm{c}^{2}\varepsilon_{\mathrm{M}}^{(2)} +\varepsilon_{\mathrm{M}}\right) ,
\end{array}
\end{equation}
cf. Example 5.2 \cite{Polizzotto2003b} in the case of gradient based methods. In the local case, the latter equation collapses to the usual form $\frac{{\sigma}}{E}   = \varepsilon_{\mathrm{M}}= \varepsilon-\alpha\Delta\theta$. Application of Eq.~(\ref{eq:EpsM}) that defines the link between strain and displacement, transforms the stress into:
\begin{equation}\label{eq:MotStress2Disp}
\begin{array}{l}
\frac{{\sigma}}{E}   = -L_\mathrm{c}^{2} {(u_0^{(3)}-w^{(4)}(z-z_0)-(\alpha \Delta\theta)^{(2)})}\\ 
+u_0^{(1)}-w^{(2)}(z-z_0)-\alpha\Delta\theta.
\end{array}
\end{equation}
In contrast to the widely employed gradient formulations, here additional constraints in the form of boundary conditions have to be imposed at $x=0$ and $x=L$ in order to assure solution of the constitutive integral model \cite{Polyanin98}:
\begin{equation}\label{eq:Stress3BC}
\begin{array}{l}
\left. L_\mathrm{c} \left( \varepsilon^{(1)}-(\alpha \Delta\theta)^{(1)}\right) - (\varepsilon-\alpha\Delta\theta)  =0 \right|_{\text{at } (0,z) } , \\
\left. L_\mathrm{c} \left( \varepsilon^{(1)}-(\alpha \Delta\theta)^{(1)}\right) + (\varepsilon-\alpha\Delta\theta) =0\right|_{\text{at } (L,z) }. \\
\end{array}
\end{equation}
These constraints assume existence and uniqueness of the solution Eq.~(\ref{eq:MotStress2}) \cite{Canadija2018}. Obviously, constraints relate kinematical quantities and imply that standard kinematical boundary conditions cannot be chosen arbitrarily. These are known as the constitutive boundary conditions and were introduced in the nonlocal beam formulation in \cite{Romano17,Romano2017c}. Consequently, the stress-driven integral formulation can be made equivalent to the gradient one if the boundary conditions Eq.~(\ref{eq:Stress3BC}) are respected.  

Alternatively, by the application of displacement and rotation, Eq.~(\ref{eq:Epsilon}), the above constraints become:
\begin{equation}\label{eq:Stress3_1}
\begin{array}{l}
L_\mathrm{c} \left( u_0^{(2)}-w^{(3)}(z-z_0) -(\alpha\Delta\theta)^{(1)}\right)  \\
- (u_0^{(1)}-w^{(2)} (z-z_0)-\alpha\Delta\theta)  =\left. 0 \right|_{x=0}, \\
L_\mathrm{c} \left( u_0^{(2)}-w^{(3)}(z-z_0) -(\alpha\Delta\theta)^{(1)}\right)  \\
+ (u_0^{(1)}-w^{(2)} (z-z_0)-\alpha\Delta\theta)  =\left. 0 \right|_{x=L}. \\
\end{array}
\end{equation}

The stress-driven model valid in a beam's point should be now homogenized for an arbitrary cross section. Hence, Eq.~(\ref{eq:MotStrain}) is multiplied by the Euler-Young's modulus (note that it must be $E\ne E(x)$ in order to do so), integrated over the cross section and finally the equilibrium equation $\int_{\Omega} \sigma \mathrm{d}A=N$ is applied. Introducing displacements Eq.~(\ref{eq:Epsilon}), this provides:
\begin{equation}\label{eq:MotStrain2}
u_0^{(1)} \int_{\Omega} E \mathrm{d} A -w^{(2)} \int_{\Omega} E(z-z_0) \mathrm{d} A-\int_{\Omega} E\alpha\Delta\theta \mathrm{d} A= \int_{0}^{L} \phi_\mathrm{c} ({x-\xi)} {N}(\xi,z) \mathrm{d} \xi,
\end{equation}
while the accompanying constitutive boundary conditions become:
\begin{equation}\label{eq:CBC_N}
\begin{array}{l}
L_\mathrm{c} \left( u_0^{(2)} \int_{\Omega} E \mathrm{d} A- w^{(3)} \int_{\Omega}E(z-z_0) \mathrm{d} A-\int_{\Omega} E(\alpha\Delta\theta)^{(1)}\mathrm{d} A\right)  \\
- (u_0^{(1)}\int_{\Omega} E\mathrm{d} A-w^{(2)} \int_{\Omega} E(z-z_0)\mathrm{d} A- \int_{\Omega}E\alpha\Delta\theta)\mathrm{d} A  =\left. 0 \right|_{x=0}, \\
L_\mathrm{c} \left( u_0^{(2)} \int_{\Omega} E \mathrm{d} A - w^{(3)} \int_{\Omega}E(z-z_0) \mathrm{d} A-\int_{\Omega} E(\alpha\Delta\theta)^{(1)}\mathrm{d} A\right)  \\
+ (u_0^{(1)}\int_{\Omega} E\mathrm{d} A-w^{(2)} \int_{\Omega} E (z-z_0)\mathrm{d} A- \int_{\Omega}E\alpha\Delta\theta)\mathrm{d} A  =\left. 0 \right|_{x=L}. \\
\end{array}
\end{equation}

The same reasoning as for the equilibrium equation in the axial direction can be now applied to bending, $\int_{\Omega} \sigma (z-z_0) \mathrm{d}A=M$. Multiplication of Eq.~(\ref{eq:MotStrain}) by $(z-z_0)$ yields:
\begin{equation}\label{eq:MotStrain3}
u_0^{(1)} \int_{\Omega} E(z-z_0) \mathrm{d} A -w^{(2)} \int_{\Omega} E(z-z_0)^2 \mathrm{d} A-\int_{\Omega} E\alpha\Delta\theta (z-z_0) \mathrm{d} A= \int_{0}^{L} \phi_\mathrm{c} ({x-\xi)} {M}(\xi,z) \mathrm{d} \xi,
\end{equation}
and
\begin{equation}\label{eq:CBC_M}
\begin{array}{l}
L_\mathrm{c} \left( u_0^{(2)} \int_{\Omega} E(z-z_0) \mathrm{d} A - w^{(3)} \int_{\Omega}E(z-z_0)^2 \mathrm{d} A-\int_{\Omega} E (\alpha\Delta\theta)^{(1)}(z-z_0)\mathrm{d} A\right)  \\
- (u_0^{(1)}\int_{\Omega} E (z-z_0) \mathrm{d} A-w^{(2)} \int_{\Omega}E (z-z_0)^2\mathrm{d} A- \int_{\Omega}E\alpha\Delta\theta(z-z_0)\mathrm{d} A ) =\left. 0 \right|_{x=0}, \\
L_\mathrm{c} \left( u_0^{(2)} \int_{\Omega} E(z-z_0) \mathrm{d} A - w^{(3)} \int_{\Omega}E(z-z_0)^2 \mathrm{d} A-\int_{\Omega} E (\alpha\Delta\theta)^{(1)}(z-z_0)\mathrm{d} A\right)  \\
+ (u_0^{(1)}\int_{\Omega} E(z-z_0) \mathrm{d} A-w^{(2)} \int_{\Omega} E (z-z_0)^2\mathrm{d} A- \int_{\Omega}E\alpha\Delta\theta(z-z_0)\mathrm{d} A )  =\left. 0 \right|_{x=L}. \\
\end{array}
\end{equation}
This constitutive model must be now placed in the strict thermodynamic framework.

\subsection{Continuum mechanics fundamentals of beam-like structures}
Having introduced the cornerstone of the stress-driven nonlocal model, the proper thermodynamic framework specialized to beams is provided. For this purpose, a brief recapitulation of balance equations from continuum mechanics is presented.  Due to one dimensional nature of the problem, vectorial and tensor quantities collapse to the scalar form. The balance of momentum in a Bernoulli-Euler beam where inertial and volume forces are deemed to be insignificant can be written as:
\begin{equation}\label{eq:BalMom}
{\sigma}^{(1)}=0.
\end{equation}
Likewise, the first law of thermodynamics is \cite{Marsden_Hughes:94}:
\begin{equation}\label{eq:1stLaw}
\rho \dot{e}+q^{(1)}={\sigma} +\rho r+\mathcal{P},
\end{equation}
while the second law is:
\begin{equation}\label{eq:2ndLaw}
\rho \dot{\eta}\ge \frac{\rho r}{\theta}-\left( \frac{q}{\theta} \right)^{(1)}.
\end{equation}
Above, $\rho(x,t)$ is the specific mass, $e(x,t)$ is the internal energy, $q(x)$ is the heat flux, $d$ is the rate of deformation, $r(x,t)$ is the heat source per unit mass,  $\theta(x,t)$ is temperature, $\mathcal{P}$ is the rate of energy supply from the neighbourhood and $\eta(x,t)$ is the specific entropy per unit mass. The origin of the rate of energy supply $\mathcal{P}$ is nonlocality. Note that for the whole beam it must be $\int_{\mathcal{B}} \mathcal{P} \mathrm{d} V=0$ in order to fulfil the balance of energy, see \cite{Polizzotto2003b} for more details. As usual, a superimposed dot represents differentiation with respect to time.  We now postulate existence of the Gibbs potential $g(\sigma,\theta)$ \cite{Malvern1969}:
\begin{equation}\label{eq:GibbsDef}
g(\sigma,\theta)=e-\eta \theta-\sigma \varepsilon.
\end{equation}
This introduces the link between the Helmholtz free energy and Gibbs potential as:
\begin{equation}\label{eq:HelmGibbs}
\psi=g+\sigma \varepsilon.
\end{equation}
Although generally valid, the latter equation does not imply that an equivalent Helmholtz function can be obtained from the Gibbs functions in all cases (and vice-versa). As demonstrated in \cite{Rajagopal2010}, such transformation is not always possible. Thus, for some type of problems, the Helmholtz free energy function is more appropriate while for other the Gibbs function is the preferred choice. 

With the Gibbs potential at hand, it can be shown that the second law of thermodynamics takes the form:
\begin{equation}\label{eq:2ndLaw-2}
-\rho \dot{g} -\dot{\sigma} {\varepsilon} - \rho \eta \dot{\theta} -  \frac{q}{\theta} \theta ^{(1)} \ge 0.
\end{equation}
Introducing the rate of the Gibbs potential as 
\begin{equation}\label{eq:HelmRate}
\dot{g}= \partial _\sigma g \dot{\sigma}+ \partial _\theta g \dot{\theta} , 
\end{equation}
it is obtained:
\begin{equation}\label{eq:2ndLaw-3}
-({\varepsilon}+\partial _\sigma g) \dot{\sigma} - \rho (\partial _\theta g +\eta) \dot{\theta} -\frac{q}{\theta} \theta ^{(1)} \ge 0.
\end{equation}
Now, standard arguments are invoked \cite{Surana2013}. Since the last equation has to hold for all arbitrary processes, it is obtained:
\begin{equation}\label{eq:2ndLaw-31}
{\varepsilon}=-\partial _\sigma g. 
\end{equation}

Finally, with Eq.~(\ref{eq:2ndLaw-31}) at hand, consider a thermoelastic process with homogeneous temperature field. Inequality (\ref{eq:2ndLaw-3}) yields the definition of entropy:
\begin{equation}\label{eq:2ndLaw-4}
\quad \eta = -\partial _\theta g.
\end{equation}
In purely thermal processes in which deformation does not take place it is: 
\begin{equation}\label{eq:2ndLawGlob}
-\frac{q}{\theta} \theta ^{(1)} \ge 0.
\end{equation}
This effectively puts constraints on the allowable direction of the heat flow. Since the present research will not address problems involving heat conduction, the last equation will not be used in the forthcoming elaborations.

\subsection{Prototype model}
Having set up the general thermodynamic framework for beams, some constitutive assumptions must be now introduced. In particular, the constitutive behaviour by means of the Gibbs function and a functional for determination of kinematical quantities is provided.

\subsubsection{Gibbs potential}
In order to put the proposed nonlocal stress-driven formulation within the thermodynamic framework described above, the Gibbs potential $g$ of the following form is used in the local formulation \cite{Lubarda04}: 
\begin{equation}\label{eq:GibbsLoc}
\begin{array}{c}
\rho g(\sigma,\theta)=-\frac{1}{2 E} \sigma^2 - \alpha \Delta \theta \sigma.
\end{array}
\end{equation}
In this particular case the above function should be modified in order to accommodate the nonlocal character of the model Eq.~(\ref{eq:MotStrain}):
\begin{equation}\label{eq:GibbsNonLoc}
\begin{array}{c}
\rho g(\sigma,\theta)=-\frac{1}{2} \sigma \int_{0}^{L} \phi_\mathrm{c} ({x-\xi)}  E^{-1} {\sigma}(\xi,z) \mathrm{d} \xi -  \sigma \alpha \Delta \theta.
\end{array}
\end{equation}
Above, the Euler-Young's modulus is assumed to be a function of the transverse coordinate, $E=E(z)$. It can be easily verified that the above choice of the Gibbs function reflects Eqs.~(\ref{eq:MotStrain}) if introduced into Eq.~(\ref{eq:2ndLaw-31}).

\subsubsection{Governing potential for beams}
To provide a variationally consistent framework, the existence of the governing potential is postulated:
\begin{equation}\label{eq:PotentialBE}
\Pi({u,w})=U-W.
\end{equation}
The internal potential $U$ is defined for the complete domain of the beam $\mathcal{B}$ for a reversible steady state thermal process ($\dot{\theta}=0$) as:
\begin{equation}\label{eq:Potential3}
\begin{array}{l}
U=\int_{\mathcal{B}} e \mathrm{d} V=\int_{\mathcal{B}} (g+\sigma \varepsilon) \mathrm{d} V= \int _0 ^t \int_{\mathcal{B}} (\partial _\sigma g \dot{\sigma} + \dot{\sigma}\varepsilon+ \sigma \dot{\varepsilon} ) \mathrm{d} V \mathrm{d} t'= \\
=\int_{\mathcal{B}} \int _0 ^\sigma (\partial _\sigma g \mathrm{d}{\sigma} + \varepsilon \mathrm{d}{\sigma}) \mathrm{d} V + \int_{\mathcal{B}} \int _0 ^\varepsilon \sigma \mathrm{d}{\varepsilon}  \mathrm{d} V = \int_{\mathcal{B}} \int _0 ^\varepsilon \sigma \mathrm{d}{\varepsilon}  \mathrm{d} V \\
\end{array}
\end{equation}
where Eq.~(\ref{eq:2ndLaw-31}) was applied. 
The potential of external forces $W$ assumes the standard form:
\begin{equation}\label{eq:Wext}
\begin{array}{l}
W=\int_L q_x {u_0} \mathrm{d} x+\int_L q_z {w} \mathrm{d} x \\ 
  + \sum_{i=0}^{L} \mathcal{N}_i {u}_{0,i}- \sum_{i=0}^{L} \mathcal{M}_i  w^{(1)}_i   \\
\end{array}
\end{equation}
where $\mathcal{N}$ are the external axial forces while $\mathcal{M}$ are bending moments at the beam end's. Note that constitutive boundary conditions Eqs.~(\ref{eq:CBC_N},\ref{eq:CBC_M}) arising from integrability constraints are indirectly included.

Finally, the solution follows from the minimum total potential energy principle:
\begin{equation}\label{eq:Solution}
\begin{array}{l}
({u_0}, {w}) = \arg \underset{{u_0}, {w}}{\inf} \Pi ({u_0}, {w}).  \\
\end{array}
\end{equation}
Required stationary conditions are discussed in derivations that follow. Finally, the variational problem at hand can be alternatively solved by the application of the minimum total complementary energy principle. 

\subsubsection{Shift of the neutral surface}
Yet remains to define the shift of the neutral surface from the centroid of the cross section. In the case of homogeneous beams, the beam's cross section rotates around the $y$ axis that is perpendicular to the plane of deformation. This axis is situated at the centroid of the cross section. It is understood that the lines parallel to the longitudinal axis laying in the neutral surface are bent, but their length does not change what results in vanishing normal stresses. However, if the beam is functionally graded, the situation is changed. Variation of beam stiffnesses can cause the shift of the neutral surface $z_0$. The section at hand will demonstrate how to determine this shift.

In the local thermoelastic problem, the displacement of any point of the beam can be considered to consists of two parts: translational and rotational part. Likewise, the axial strain of mechanical origin in a longitudinal cross section defined by the coordinate $z$ is composed of three parts, Eq.~(\ref{eq:EpsM}). Translation is related to the uniform elongation term $u_0^{(1)}$ and thermal elongation $\alpha \Delta \theta$, while the rotation due to bending causes $-w ^{(2)}(z-z_0)$. In the neutral surface, axial displacement due to rotation (and consequently strain due to bending) must vanish. This can be used to determine the shift $z_0$. Now, the stresses in an arbitrary cross section $x$ are calculated as:
\begin{equation}\label{eq:ShiftStress}
\sigma(x,z)= E \varepsilon_{\mathrm{M}} =E (u_0^{(1)} -w ^{(2)}(z-z_0)-\alpha \Delta \theta).
\end{equation}
Since in the neutral surface stresses (and strains) due to rotation of the cross section should vanish it is obtained:
\begin{equation}\label{eq:ShiftStress1}
\begin{array}{l}
0=-Ew ^{(2)}(z-z_0).
\end{array}
\end{equation}
Now, integration of the above equation over the cross section provides the shift $z_0$:
\begin{equation}\label{eq:z0}
\begin{array}{c}
\int_\Omega (z-z_0) E \mathrm{d} A = 0 \\ 
\int_\Omega z E \mathrm{d} A -\int_\Omega z_0 E \mathrm{d} A =0 \quad \Rightarrow \quad z_0=\frac{\int_\Omega z E  \mathrm{d} A }{\int_\Omega E \mathrm{d} A }.
\end{array}
\end{equation}

Note that in the case of $E=E(x,z)$, the above shift becomes a function of the axial coordinate, $z_0=z_0(x)$ significantly complicating the formulation. However, in some special cases simpler solutions can be obtained. For instance, let the Euler-Young's modulus can be expressed as $E(x,z)=E_x(x)E_z(z)$.  The advantage of such assumption is obvious when $E$ is introduced into Eq.~(\ref{eq:z0}):
\begin{equation}\label{eq:z0-3}
z_0=\frac{\int_\Omega z E_x E_z  \mathrm{d} A }{\int_\Omega E_x E_z \mathrm{d} A }=\frac{E_x \int_\Omega z  E_z  \mathrm{d} A }{E_x \int_\Omega E_z \mathrm{d} A }=\frac{\int_\Omega z  E_z  \mathrm{d} A }{\int_\Omega E_z \mathrm{d} A },
\end{equation}
conveniently transforming $z_0$ to a constant. Obviously, the identical equation is obtained for the case $E=E(z)=E_z$.

If a beam is homogeneous ($E=\text{const.}$) or $E=E(x)=E_x$, the shift of the neutral surface $z_0$ is zero, since Eq.~(\ref{eq:z0}) collapses to
\begin{equation}\label{eq:LF-A1}
z_0=\frac{\int_\Omega z \mathrm{d} A }{A}=0.
\end{equation}
Note that the first moment of area still vanishes in the FG formulation, $\int_\Omega z \mathrm{d} A=0$.

If the nonlocal formulation, the starting counterpart is Eq.~(\ref{eq:MotStrain}). Suppose again that  $E(x,z)=E_x(x)E_z(z)$. This gives:
\begin{equation}\label{eq:ShiftNL1}
E_z(u_0^{(1)} -w ^{(2)}(z-z_0)-\alpha \Delta \theta) = \int_{0}^{L} \phi_\mathrm{c} ({x-\xi)}  E_x^{-1} {\sigma}(\xi,z) \mathrm{d} \xi.
\end{equation}
Following the same reasoning as in the local formulation, at the neutral surface strain contribution from the rotation term must vanish, again providing:
\begin{equation}\label{eq:ShiftNL2}
- E_z w ^{(2)}(z-z_0)=0 \quad \Rightarrow \quad z_0=\frac{\int_\Omega z E_z  \mathrm{d} A }{\int_\Omega E_z \mathrm{d} A },
\end{equation}
repeating the result from the local case. 
 
\subsection{Stationarity conditions}
\subsubsection{Stationarity with respect to $\delta_{{u}_0} \Pi$}
This section deals with the stationary conditions that provide governing equations and boundary conditions. First term $\delta_{{u}_0} \Pi=0$ is obtained as follows. Starting from the total energy potential Eq.~(\ref{eq:PotentialBE}) and strain Eq.~(\ref{eq:Epsilon}), it is obtained: 
\begin{equation}\label{eq:U_u0-1}
\delta_{{u}_0} \Pi = \int_{\mathcal{B}} \sigma  \delta {u}_0^{(1)} \; \mathrm{d} V-\int_L q_x \delta {u_0} \mathrm{d} x-\sum_{i=0}^{L} \mathcal{N}_i \delta {u}_{0,i}=0.
\end{equation}
Integration by parts with respect to the longitudinal coordinate $x$ provides:
\begin{equation}\label{eq:U_u0-2}
\delta_{{u}_0} \Pi = \int_{\Omega} \sigma \; \mathrm{d} A \delta {u}_0 \left| _0 ^L \right. - \int_{\mathcal{B}} \sigma^{(1)} \delta {u}_0 \; \mathrm{d} V -\int_L q_x \delta {u_0} \mathrm{d} x-\sum_{i=0}^{L} \mathcal{N}_i \delta {u}_{0,i}=0.
\end{equation}
The first term will provide boundary conditions as
\begin{equation}\label{eq:U_u0-3}
\begin{array}{l}
\left. \int_{\Omega} \sigma \; \mathrm{d} A =-\mathcal{N} \right| _{x=0 } \quad \text{ or prescribe } u_0(0), \\
\left. \int_{\Omega} \sigma \; \mathrm{d} A = \mathcal{N} \right| _{x=L } \quad \text{ or prescribe } u_0(L).
\end{array}
\end{equation}
It is emphasized that constitutive boundary conditions Eq.~(\ref{eq:CBC_N}) accompany above listed standard boundary conditions.

The second term in Eq.~(\ref{eq:U_u0-2}) gives in the same manner:
\begin{equation}\label{eq:U_u0-8}
-\int_{\mathcal{B}} \sigma^{(1)}  \delta {u}_0 \mathrm{d} V-\int_L q_x \delta {u_0} \mathrm{d} x=0,
\end{equation}
or having in mind arbitrariness of the virtual axial displacement:
\begin{equation}\label{eq:U_u0-8b}
\int_{\Omega} \sigma^{(1)} \mathrm{d} A+ q_x =0.
\end{equation}

Due to Eq.~(\ref{eq:MotStress2DispEpsM}), $\sigma^{(1)} $ is:
\begin{equation}\label{eq:U_u0-8a}
\sigma^{(1)}= E\left( -L_\mathrm{c}^{2}\varepsilon_{\mathrm{M}}^{(3)} +\varepsilon_{\mathrm{M}}^{(1)}\right).
\end{equation}
Explicit forms for $\varepsilon_{\mathrm{M}}^{(1)}$ and $\varepsilon_{\mathrm{M}}^{(3)}$ will be introduced later. Alternatively, Eq.~(\ref{eq:U_u0-8b}) can be also rewritten  as:
\begin{equation}\label{eq:U_u0-8c}
N^{(1)} + q_x =0
\end{equation}
if the equilibrium equation in the axial direction is employed.

\subsubsection{Stationarity with respect to $\delta_{{w}} \Pi$} \label{sec:wStat}
Second stationary condition $\delta_{{w}} \Pi=0$ gives:
\begin{equation}\label{eq:U_phi-1}
\begin{array}{l}
\delta_{{w}} \Pi=-\int_{\mathcal{B}} \sigma (z-z_0)  \delta {w} ^{(2)}\; \mathrm{d} V  -\int_L q_z \delta {w} \mathrm{d} x + \sum_{i=0}^{L}  \mathcal{M}_i  \delta w^{(1)}_i=0.
\end{array}
\end{equation}

The same steps as in the first stationary condition are followed. Integrating by parts twice yields:
\begin{equation}\label{eq:U_phi-2}
\begin{array}{l}
\delta_{{w}} \Pi = -\int_{\Omega} \sigma (z-z_0) \delta {w}^{(1)}  \; \mathrm{d} A \left| _0 ^L \right. +\int_{\Omega} \sigma ^{(1)} (z-z_0)  \mathrm{d} A \delta {w} \left| _0 ^L \right. \\
-\int_{\mathcal{B}} \sigma^{(2)}(z-z_0)\delta {w} \; \mathrm{d} V  -\int_L q_z \delta {w} \mathrm{d} x + \sum_{i=0}^{L}  \mathcal{M}_i  \delta w^{(1)}_i=0.
\end{array}
\end{equation}
The first two terms with Eq.~(\ref{eq:MotStress2Disp}) again provide boundary conditions. The first boundary condition exploits arbitrariness of $\delta w^{(1)}$ at $x\in \lbrace 0,L \rbrace$:
\begin{equation}\label{eq:U_u0-31}
\begin{array}{l}
\left. \int_{\Omega} \sigma (z-z_0) \; \mathrm{d} A = -\mathcal{M} \right| _{x =0 } \text{ or prescribe } w^{(1)}(0),  \\
\left. \int_{\Omega} \sigma (z-z_0) \; \mathrm{d} A = \mathcal{M} \right| _{x =L } \text{ or prescribe }  w^{(1)}(L)  
\end{array}
\end{equation}
and the second one $\delta w$ at $x\in \lbrace 0,L \rbrace$:
\begin{equation}\label{eq:U_u0-32}
\begin{array}{l}
\left. \int_{\Omega}  \sigma^{(1)} (z-z_0)   \mathrm{d} A = 0 \right| _{x=0 } \quad \text{ or prescribe } w(0), \\
\left. \int_{\Omega}  \sigma^{(1)} (z-z_0)  \mathrm{d} A = 0 \right| _{x=L } \quad \text{ or prescribe } w(L).
\end{array}
\end{equation}
The constitutive boundary conditions Eq.~(\ref{eq:CBC_M}) arising from the integrability conditions must be included as well.

The third and fourth term in Eq.~(\ref{eq:U_phi-2}) gives:
\begin{equation}\label{eq:U_u0-34}
\begin{array}{l}
-\int_{\mathcal{B}} \sigma^{(2)} (z-z_0)  \delta {w} \mathrm{d} V -\int_L q_z \delta {w} \mathrm{d} x =0,
\end{array}
\end{equation}
or
\begin{equation}\label{eq:U_u0-35}
\begin{array}{l}
-\int_{\Omega}\sigma^{(2)} (z-z_0)   \mathrm{d} A  - q_z =0. \\
\end{array}
\end{equation}
The first derivative of stress is defined by Eq.~(\ref{eq:U_u0-8a}), while the second one is:
\begin{equation}\label{eq:Stress2}
\sigma^{(2)}= E\left(- L_\mathrm{c}^{2}\varepsilon_{\mathrm{M}}^{(4)} +\varepsilon_{\mathrm{M}}^{(2)}\right).
\end{equation}
Like in the axial case, Eq.~(\ref{eq:U_u0-35}) can be transformed by means of the equilibrium equation into:
\begin{equation}\label{eq:U_u0-35a}
\begin{array}{l}
M^{(2)} + q_z =0. \\
\end{array}
\end{equation}

\section{Beam displacements of nonlocal FG beams}
Equations defining beam displacements are derived for functional grading in one direction of the Euler-Young's modulus $E=E(z)$. The coefficient of thermal expansion and temperature are of a general form: $\alpha=\alpha(x,z)$, $\theta=\theta(x,z)$. At the same time, following notation is introduced:
\begin{equation}\label{eq:Notation_NLF_Ez-1}
\begin{array}{l}
\Phi^\mathrm{N}_{m}(x)=\int_\Omega  E (\alpha \Delta\theta)^{(m)} \mathrm{d} A, \\
\Phi^\mathrm{M}_{m}(x)=\int_\Omega  E (\alpha \Delta\theta)^{(m)} (z-z_0) \mathrm{d} A ,\\
k^\mathrm{AE}(x)=\int_\Omega  E \mathrm{d} A , \\
k^\mathrm{SE}(x)=\int_\Omega  E(z-z_0) \mathrm{d} A,  \\
k^\mathrm{IE}(x)=\int_\Omega  E(z-z_0)^2 \mathrm{d} A. \\
\end{array}
\end{equation}
In the present case, the shift of the neutral surface is not zero. The first and second derivative of the stress $\sigma^{(1)},\sigma^{(2)} $ now have nonlocal character, Eq.~(\ref{eq:MotStress2}):
\begin{equation}\label{eq:U_u0-NLF_1}
\begin{array}{l}
\sigma^{(1)}= -E L_\mathrm{c}^{2} (u_0^{(4)} -w ^{(5)}(z-z_0)-(\alpha \Delta \theta)^{(3)}) + E(u_0^{(2)} -w ^{(3)}(z-z_0)-(\alpha \Delta \theta)^{(1)}), \\
\sigma^{(2)}= -E L_\mathrm{c}^{2} (u_0^{(5)} -w ^{(6)}(z-z_0)-(\alpha \Delta \theta)^{(4)}) + E(u_0^{(3)} -w ^{(4)}(z-z_0)-(\alpha \Delta \theta)^{(2)}). \\
\end{array}
\end{equation}
With above result at hand, the governing equation for the axial displacement follows from Eq.~(\ref{eq:U_u0-8b}) as
\begin{equation}\label{eq:U_u0-NLF_3}
\begin{array}{l}
\int_{\Omega} \left(-E L_\mathrm{c}^{2} (u_0^{(4)} -(\alpha \Delta \theta)^{(3)}) + E(u_0^{(2)} -(\alpha \Delta \theta)^{(1)}) \right) \mathrm{d} A+ q_x  =0, \\
\end{array}
\end{equation}
where some terms were cancelled due to the property of the neutral surface. With the notation Eq.~(\ref{eq:Notation_NLF_Ez-1}), the equation describing axial deformation is:
\begin{equation}\label{eq:U_u0-NLF_Ez1}
\begin{array}{l}
-L_\mathrm{c}^{2} u_0^{(4)} k^\mathrm{AE}  +L_\mathrm{c}^{2} \Phi^\mathrm{N}_{3} +u_0^{(2)} k^\mathrm{AE} -\Phi^\mathrm{N}_{1} + q_x  =0. \\
\end{array}
\end{equation}
Boundary conditions again follows from Eq.~(\ref{eq:U_u0-3}) as:
\begin{equation}\label{eq:U_u0-NLF_6}
\begin{array}{l}
\left. -L_\mathrm{c}^{2} u_0^{(3)}k^\mathrm{AE} +L_\mathrm{c}^{2}\Phi^\mathrm{N}_{2}  + u_0^{(1)} k^\mathrm{AE}  -\Phi^\mathrm{N}_{0}  =-\mathcal{N} \right| _{x = 0} \quad \text{ or prescribe } u_0(0),  \\
\left. -L_\mathrm{c}^{2} u_0^{(3)}k^\mathrm{AE} +L_\mathrm{c}^{2}\Phi^\mathrm{N}_{2}  + u_0^{(1)} k^\mathrm{AE}  -\Phi^\mathrm{N}_{0}  = \mathcal{N} \right| _{x=L} \quad \text{ or prescribe } u_0(L) \\
\end{array}\end{equation}
and from Eq.~(\ref{eq:CBC_N}):
\begin{equation}\label{U_u0-NLF_6a}
\begin{array}{l}
L_\mathrm{c} \left( u_0^{(2)} \int_{\Omega} E \mathrm{d} A-\int_{\Omega} E(\alpha\Delta\theta)^{(1)}\mathrm{d} A\right)  - (u_0^{(1)}\int_{\Omega} E\mathrm{d} A- \int_{\Omega}E\alpha\Delta\theta)\mathrm{d} A  =\left. 0 \right|_{x=0}, \\
L_\mathrm{c} \left( u_0^{(2)} \int_{\Omega} E \mathrm{d} A -\int_{\Omega} E(\alpha\Delta\theta)^{(1)}\mathrm{d} A\right)  + (u_0^{(1)}\int_{\Omega} E\mathrm{d} A- \int_{\Omega}E\alpha\Delta\theta)\mathrm{d} A  =\left. 0 \right|_{x=L}. \\
\end{array}
\end{equation}

The governing equation for bending is derived from Eq.~(\ref{eq:U_u0-35}):
\begin{equation}\label{eq:U_w_NLF_7}
\begin{array}{l}
-\int_{\Omega} \left( \sigma^{(2)} (z-z_0) \right) \mathrm{d} A  - q_z =0. \\
\end{array}
\end{equation}
Introduction of the second derivative of stress, Eq.~(\ref{eq:U_u0-NLF_1})$_2$ and accounting for the neutral surface effects, it is obtained:
\begin{equation}\label{eq:U_w_NLF_15}
\begin{array}{l}
L_\mathrm{c}^{2} (-w ^{(6)} k^\mathrm{IE}  -\Phi^\mathrm{M}_{4}) +w ^{(4)} k^\mathrm{IE}+\Phi^\mathrm{M}_{2}   - q_z  =0, \\
\end{array}
\end{equation}
cf. Remark 7.1 in \cite{Romano2017c} for the isothermal case. Thus, in the contrast to the standard fourth order differential equation, the present sixth order case includes constitutive boundary conditions Eq.~(\ref{eq:U_w_NLF_Ez6}) into the formulation.

The boundary conditions are defined by Eqs.~(\ref{eq:U_u0-31}, \ref{eq:U_u0-32}):
\begin{equation}\label{eq:U_u0-313}
\begin{array}{l}
\left. -L_\mathrm{c}^{2} (w ^{(4)} k^\mathrm{IE} + \Phi^\mathrm{M}_{2}) +w ^{(2)} k^\mathrm{IE}+\Phi^\mathrm{M}_{0} =-\mathcal{M} \right| _{x=0} \text{ or prescribe } w^{(1)}(0), \\
\left. -L_\mathrm{c}^{2} (w ^{(4)} k^\mathrm{IE} + \Phi^\mathrm{M}_{2}) +w ^{(2)} k^\mathrm{IE}+\Phi^\mathrm{M}_{0} = \mathcal{M} \right| _{x=L} \text{ or prescribe } w^{(1)}(L) \\
\end{array}
\end{equation}
and the second one:
\begin{equation}\label{eq:U_w_NLF_Ez5}
\begin{array}{l}
\left. L_\mathrm{c}^{2} (w ^{(5)} k^\mathrm{IE}+ \Phi^\mathrm{M}_{3}) -w ^{(3)} k^\mathrm{IE} -\Phi^\mathrm{M}_{1} =0 \right| _{x \in \left\lbrace  0,L \right\rbrace } \text{ or prescribe } w(0), w(L), \\
\end{array}
\end{equation}
along with conditions Eq.~(\ref{eq:CBC_M})
\begin{equation}\label{eq:U_w_NLF_Ez6}
\begin{array}{l}
L_\mathrm{c} \left( - w^{(3)} \int_{\Omega}E(z-z_0)^2 \mathrm{d} A-\int_{\Omega} E (\alpha\Delta\theta)^{(1)}(z-z_0)\mathrm{d} A\right)  \\
- (-w^{(2)} \int_{\Omega} E (z-z_0)^2\mathrm{d} A- \int_{\Omega}E\alpha\Delta\theta(z-z_0)\mathrm{d} A ) =\left. 0 \right|_{x=0}, \\
L_\mathrm{c} \left( - w^{(3)} \int_{\Omega}E(z-z_0)^2 \mathrm{d} A-\int_{\Omega} E (\alpha\Delta\theta)^{(1)}(z-z_0)\mathrm{d} A\right)  \\
+ (-w^{(2)} \int_{\Omega} E (z-z_0)^2\mathrm{d} A- \int_{\Omega}E\alpha\Delta\theta(z-z_0)\mathrm{d} A )  =\left. 0 \right|_{x=L}. \\
\end{array}
\end{equation}
With above differential equations and accompanying boundary conditions the formulation is completely defined.

\section{Examples}
\label{Examples}
\subsection{Homogeneous doubly clamped bar heated by the uniform thermal field}
The introductory example deals with a nonlocal doubly clamped bar of length $L$ subjected to the uniform temperature field $\Delta \theta$. The similar example was analysed in \cite{Canadija2018} (Example 5.2), but with formulation that is derived slightly differently than in the present example. In particular, the approach chosen in the latter paper uses differential equation of third order for describing axial displacements, while here fourth order differential equation is employed. Naturally, this requires different number of boundary conditions. The bar is not functionally graded and material properties are defined by constant $E, \alpha$. The cross section is rectangular with dimensions $b$ and $h$.

Thus, the problem that should be solved is governed by Eq.~(\ref{eq:U_u0-NLF_Ez1}):
\begin{equation}\label{eq:Ex1-1case1}
\begin{array}{l}
-L_\mathrm{c}^{2} u_0^{(4)} AE +u_0^{(2)} AE -\Phi^\mathrm{N}_{1}=0. \\
\end{array}
\end{equation}
The constitutive boundary conditions are:
\begin{equation}\label{eq:Ex1-2case1}
\begin{array}{l}
\left. L_\mathrm{c} \left( u_0^{(2)} -(\alpha\Delta\theta)^{(1)} \right)  - (u_0^{(1)}-\alpha \Delta\theta)  =0  \right|_ {x=0},\\
\left. L_\mathrm{c} \left( u_0^{(2)} -(\alpha\Delta\theta)^{(1)} \right)  + (u_0^{(1)}-\alpha \Delta\theta)  =0  \right|_ {x=L}.\\
\end{array}
\end{equation}
At the left end, $u_0(0)=0$ was prescribed. The boundary condition at the right end follows from Eq.~(\ref{eq:U_u0-NLF_6}):
\begin{equation}\label{eq:Ex1-3case1}
\begin{array}{l}
\left. -L_\mathrm{c}^{2} u_0^{(3)}k^\mathrm{AE} + u_0^{(1)} k^\mathrm{AE} -\Phi^\mathrm{N}_{0}  =P \right| _{x=L},
\end{array}
\end{equation}
where $P$ denotes the unknown axial reaction in the support. The solution of the problem coincides with the one in \cite{Canadija2018}:
\begin{equation}\label{eq:Ex1-4case1}
u_0(x)=\alpha \Delta \theta x+\frac{ P}{EA }x- L_c \frac{ P }{2 EA }    \left(e^{\frac{x}{L_c}}-1\right) \left(e^{\frac{-L}{L_c}}+e^{\frac{-x}{L_c}}\right).
\end{equation}
The unknown support reaction follows from the remaining boundary condition $u_0(L)=0$ as:
\begin{equation}\label{eq:Ex1-4case1a}
0=\alpha \Delta \theta L+\frac{ P}{EA }L- L_c \frac{ P }{EA }    \left(e^{\frac{L}{L_c}}-1\right) e^{\frac{-L}{L_c}}.
\end{equation}
or after straightforward algebraic manipulation:
\begin{equation}\label{eq:Ex1-5case1}
P = -\frac{\alpha \Delta \theta A E L  e^{L/ L_c}}{e^{L/ L_c} (L- L_c)+ L_c}.
\end{equation}
This confirms the equivalence of both approaches.

\subsection{Axial thermal deformation of a homogeneous doubly clamped bar heated by the longitudinally varying temperature field}
The second example deals with a nonlocal doubly clamped bar of length $2L$ subjected to the varying temperature field $\Delta \theta=-\theta_0 \left( 1-\frac{x^2}{L^2} \right)$. The origin of the coordinate system is positioned at the midpoint of bar, so the endpoints are situated at $x=-L$ and $x=L$. Thus, $L_c=2 L \; c$. The present example is chosen in order to compare the solution of the strain gradient model presented in \cite{Polizzotto2003b}, Example 5.2 to the integral stress-driven model introduced in this manuscript. Material properties and cross section dimensions are similar like in the previous example.

Governing differential equation is again defined by Eq.~(\ref{eq:U_u0-NLF_Ez1}):
\begin{equation}\label{eq:Ex1-1}
\begin{array}{l}
-L_\mathrm{c}^{2} u_0^{(4)} AE +u_0^{(2)} AE -\Phi^\mathrm{N}_{1}=0, \\
\end{array}
\end{equation}
where $\Phi^\mathrm{N}_{3}=0$, $\Phi^\mathrm{N}_{1}=2 \alpha E \theta_0 \frac{A}{L^2} x$ and $q_x=0$ were used. The governing equation is the same for the stress-driven formulation and the gradient one. The general solution of the equation is:
\begin{equation}\label{eq:Ex1-2}
\begin{array}{l}
u_0(x)=\frac{\alpha \theta_0 x^3}{3 L^2}+ L_c^2 e^{x/L_c} C_1 + L_c^2 e^{-x/L_c} C_2+x C_3 +C_4,
\end{array}
\end{equation}
where $C_i, i=1,2,3,4$ are integration constants. To obtain these, boundary conditions must be enforced. In both formulations, displacements at the end of the bar must be zero:
\begin{equation}\label{eq:Ex1-3}
\begin{array}{l}
u_0(-L)=u_0(L)=0.
\end{array}
\end{equation}
The rest of boundary conditions are different in each formulation. In the gradient one these are:
\begin{equation}\label{eq:Ex1-4}
\begin{array}{l}
u_0^{(2)}(-L)=u_0^{(2)}(L)=0.
\end{array}
\end{equation}
In the stress-driven integral formulation, the boundary conditions that must be fulfilled follow from Eq.~(\ref{eq:CBC_N})
\begin{equation}\label{eq:Ex1-5}
\begin{array}{l}
\left. L_\mathrm{c} \left( u_0^{(2)} -(\alpha\Delta\theta)^{(1)} \right)  - (u_0^{(1)}-\alpha \Delta\theta)  =0  \right|_ {x=-L},\\
\left. L_\mathrm{c} \left( u_0^{(2)} -(\alpha\Delta\theta)^{(1)} \right)  + (u_0^{(1)}-\alpha \Delta\theta)  =0  \right|_ {x=L}.\\
\end{array}
\end{equation}

\begin{figure}
	\centering
	\includegraphics[scale=0.45]{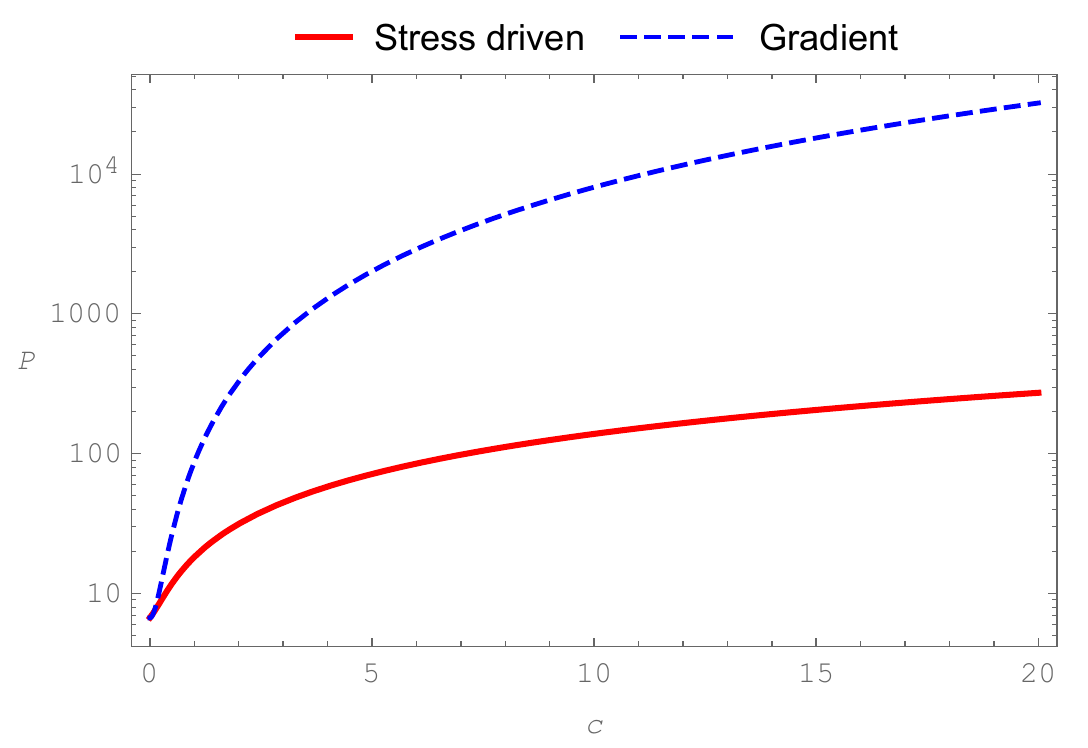}
	\caption{Dependence of support reactions on the small size parameter for the integral and gradient approach} 
	\label{fig:Ex1_3_P}
\end{figure}

The gradient formulation provides axial displacement as:
\begin{equation}\label{eq:Ex1-6}
\begin{array}{l}
u_0(x)=\frac{1}{3} \alpha \theta_0 \left( \frac{x^3}{L^2} -x + \frac{-6 L L_c^2 \mathrm{csch} \left(\frac{L}{L_c}\right) \sinh \left(\frac{x}{L_c}\right)+6 L_c^2 x}{L^2}\right),
\end{array}
\end{equation}
while the integral one gives:
\begin{equation}\label{eq:Ex1-7}
\begin{array}{l}
 u_0(x)=\frac{1}{3} \alpha \theta_0 \left(\frac{x^3}{L^2}- x +\frac{2 L_c \left(L \sinh \left(\frac{x}{L_c}\right)-x \sinh \left(\frac{L}{L_c}\right)\right)}{L_c \sinh \left(\frac{L}{L_c}\right)-L e^{L/L_c}}\right).
\end{array}
\end{equation}
Support reactions are determined as $\left. P=\int_{\Omega} \sigma \mathrm{d}A \right|_{x=L}$:
\begin{equation}\label{eq:Ex1-8}
\begin{array}{c}
P=\frac{2 \alpha A E \theta_0 e^{1/2 c}  }{3 e^{1/2 c} -6 c  \sinh \left(\frac{1}{2 c}\right)} \quad \text{(integral)}, \\
P=\frac{2}{3} \alpha A E \theta_0 \left(12 c ^2 +1\right)\quad \text{(gradient)}.
\end{array}
\end{equation}
All results are graphically presented for $L=1, \alpha=0.1, \theta_0=10, c=0.2$, $b=1$, $h=1$. Dependence of support reactions on the small size parameter is presented in the logarithmic plot in Fig.~\ref{fig:Ex1_3_P}. Obviously, the gradient approach is governed by the quadratic function and results in much higher reaction forces than the integral one as the small size parameter is increased. The integral formulation results in a more complex dependence, most easily graphically interpreted. This formulation shows much slower increase than the gradient method reaction.

\begin{figure}
	\centering
	\includegraphics[scale=0.45]{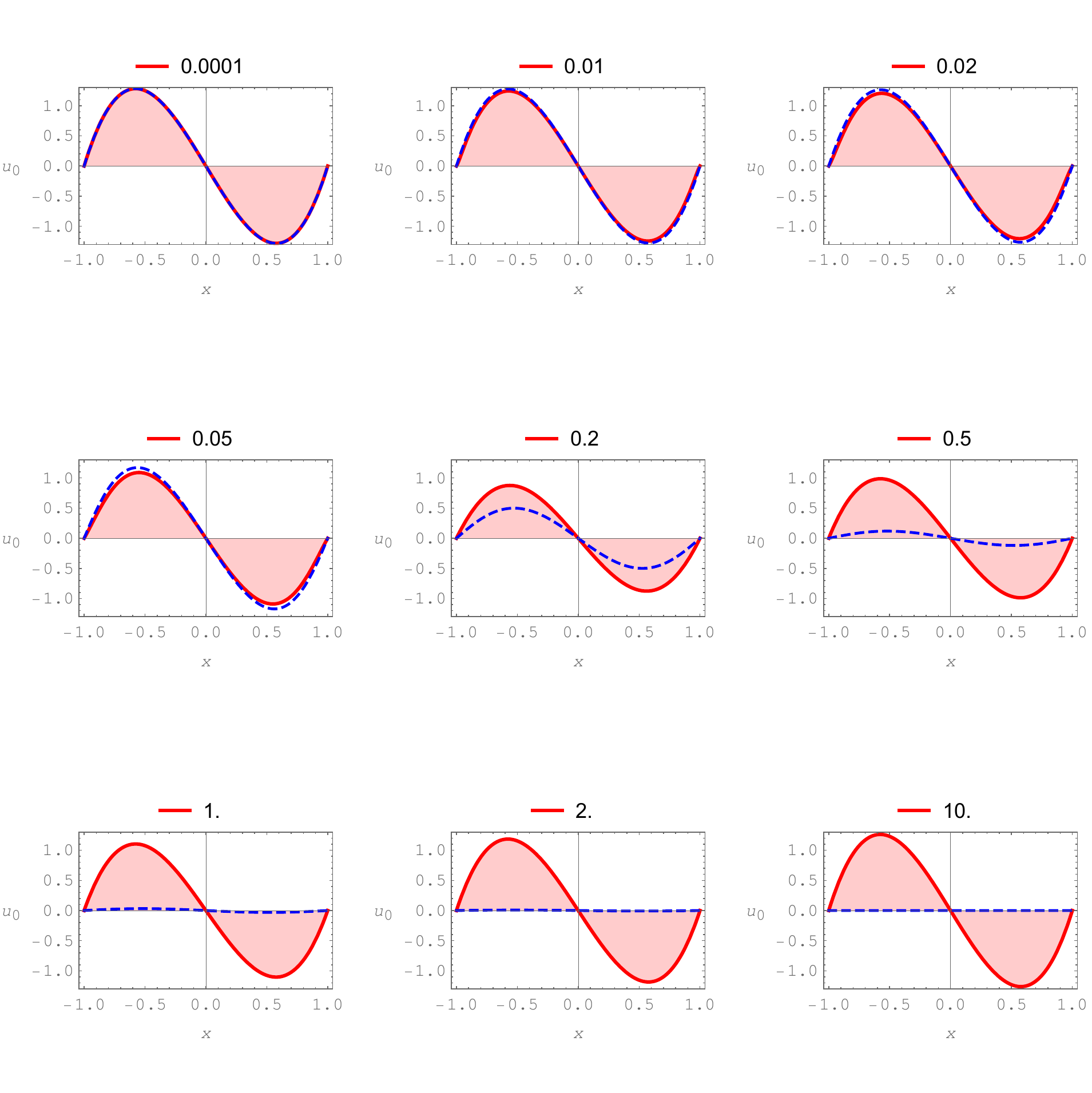}
	\caption{Distribution of the axial displacement $u_0$ along the thermally loaded doubly clamped beam for $c \in \left\lbrace 0.0001, 0.01, 0.02, 0.05, 0.2, 0.5, 1.0, 2.0, 10\right\rbrace$. Red continuous line - integral, blue dashed line - gradient formulation} 
	\label{fig:Ex1_1_u}
\end{figure}

\begin{figure}
	\centering
	\includegraphics[scale=0.45]{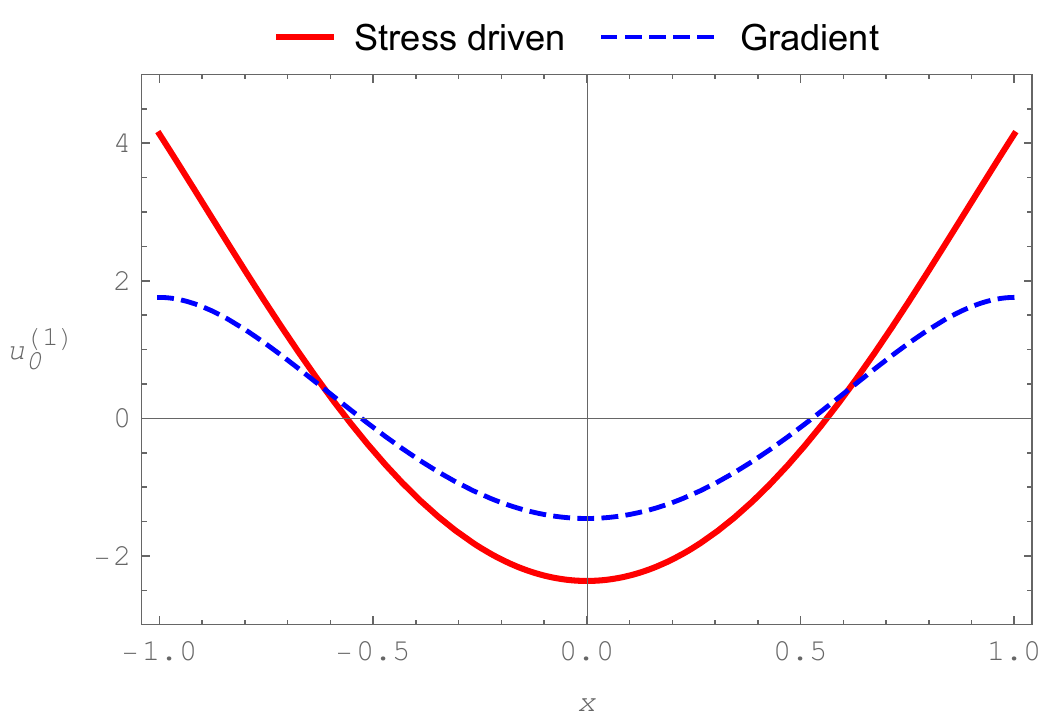}
	\caption{Distribution of $u_0^{(1)}=\varepsilon$ along the thermally loaded doubly clamped beam, $c=0.2$} 
	\label{fig:Ex1_2a_u1}
\end{figure}

\begin{figure}
	\centering
	\includegraphics[scale=0.45]{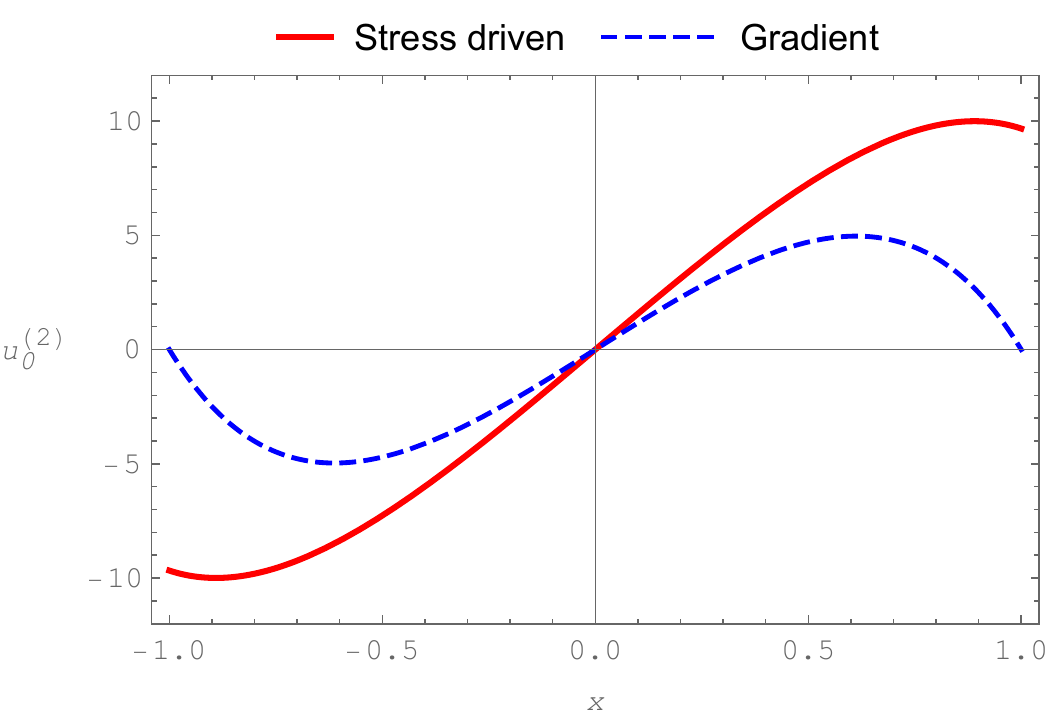}
	\caption{Distribution of $u_0^{(2)}$ along the thermally loaded doubly clamped beam, $c=0.2$} 
	\label{fig:Ex1_3_u2}
\end{figure}

Solutions for axial displacements also differ. Distribution of the axial displacement $u_0$ shows almost identical solutions for the smallest values of $c$, Fig.~\ref{fig:Ex1_1_u}. As the small size parameter is increased, solutions obtained with the gradient method show monotonic increase in beam's stiffness. Displacements at the beam ends are obviously identical due to boundary conditions Eq.~(\ref{eq:Ex1-3}) for both formulations. Unlike the gradient solution, the integral approach initially exhibits stiffening, but after a certain threshold the displacements again increase. This can be explained by the different last terms in parentheses of Eqs.~(\ref{eq:Ex1-6}, \ref{eq:Ex1-7}). Such behaviour was already noticed earlier, see \cite{Canadija2018} for further clarifications. The strain distribution, Fig.~\ref{fig:Ex1_2a_u1} also manifests differences between two methods. The origin of such behaviour is the difference in boundary conditions Eqs.~(\ref{eq:Ex1-4}, \ref{eq:Ex1-5}). Since the gradient approach enforces the second derivative of the axial displacement to be zero at the beam ends (Fig.~\ref{fig:Ex1_3_u2}), this has the consequence that the tangent to curve describing the strain distribution must be horizontal at ends. This is not the case in the integral formulation.

\subsection{Thermally loaded FG cantilever beam}
\label{sec:Example3}
To demonstrate dependency of results on the neutral surface shift, a cantilever FG beam is considered. Temperature field $\Delta \theta=\theta_1 z^2$ is imposed on the beam. At $x=L$ the beam is loaded by the axial force $P$ and couple $M$. The coefficient of thermal expansion is $\alpha$, the Euler-Young's modulus is linearly distributed across the height of the beam $E(z)=E_0+E_1(z+h/2)$, while the beam rectangular cross section is defined by the height $h$ and the width $b$ as usual. 

As a consequence of the functional grading, the neutral surface is shifted by (Eq.~(\ref{eq:ShiftNL2})):
\begin{equation}\label{eq:Ex3_1}
 z_0=\frac{\int_\Omega z E_z  \mathrm{d} A }{\int_\Omega E_z \mathrm{d} A }= \frac{2E_1 I_y}{2E_0 A- b h^2 E_1}.
\end{equation}

The problem is defined by the set of differential equations Eqs.~(\ref{eq:U_u0-NLF_Ez1}, \ref{eq:U_w_NLF_15}):
\begin{equation}\label{eq:Ex3_2}
\begin{array}{l}
-L_\mathrm{c}^{2} u_0^{(4)} k^\mathrm{AE} +u_0^{(2)} k^\mathrm{AE} =0, \\
-L_\mathrm{c}^{2} w ^{(6)}  k^\mathrm{IE} +w ^{(4)}  k^\mathrm{IE} =0 \\
\end{array}
\end{equation}
where terms $\Phi^\mathrm{N}_{1}$, $\Phi^\mathrm{N}_{3}$, $\Phi^\mathrm{M}_{2}$, $\Phi^\mathrm{M}_{4}$ vanish since the temperature does not depend on the longitudinal coordinate $x$. The set of boundary conditions follows from Eqs.~(\ref{eq:U_u0-NLF_6},\ref{U_u0-NLF_6a},\ref{eq:U_u0-313},\ref{eq:U_w_NLF_Ez5},\ref{eq:U_w_NLF_Ez6}):
\begin{equation}\label{eq:Ex3_3}
\begin{array}{l}
u_0(0)=0, w(0)=0, w^{(1)}=0, \\
\left. -L_\mathrm{c}^{2} u_0^{(3)}k^\mathrm{AE} + u_0^{(1)} k^\mathrm{AE}  -\Phi^\mathrm{N}_{0}  =P \right| _{x=L},  \\
\left. -L_\mathrm{c}^{2} w ^{(4)} k^\mathrm{IE} + w ^{(2)} k^\mathrm{IE}+\Phi^\mathrm{M}_{0} = M \right| _{x=L}, \\
\left. L_\mathrm{c}^{2} w ^{(5)} k^\mathrm{IE}  -w ^{(3)} k^\mathrm{IE}  =0 \right| _{x=L }, \\
L_\mathrm{c}  u_0^{(2)} \int_{\Omega} E \mathrm{d} A  - (u_0^{(1)}\int_{\Omega} E\mathrm{d} A- \int_{\Omega}E\alpha\Delta\theta)\mathrm{d} A  =\left. 0 \right|_{x=0}, \\
L_\mathrm{c}  u_0^{(2)} \int_{\Omega} E \mathrm{d} A  + (u_0^{(1)}\int_{\Omega} E\mathrm{d} A- \int_{\Omega}E\alpha\Delta\theta)\mathrm{d} A  =\left. 0 \right|_{x=L}, \\
-L_\mathrm{c} w^{(3)} \int_{\Omega}E(z-z_0)^2 \mathrm{d} A - (-w^{(2)} \int_{\Omega} E (z-z_0)^2\mathrm{d} A- \int_{\Omega}E\alpha\Delta\theta(z-z_0)\mathrm{d} A ) =\left. 0 \right|_{x=0}, \\
-L_\mathrm{c} w^{(3)} \int_{\Omega}E(z-z_0)^2 \mathrm{d} A + (-w^{(2)} \int_{\Omega} E (z-z_0)^2\mathrm{d} A- \int_{\Omega}E\alpha\Delta\theta(z-z_0)\mathrm{d} A )  =\left. 0 \right|_{x=L}, \\
\end{array}
\end{equation}
where $\Phi^\mathrm{N}_{2}$, $\Phi^\mathrm{M}_{3}$, $\Phi^\mathrm{M}_{1}$ again disappear.

Solutions of the above problem are:
\begin{equation}\label{eq:Ex3_4}
\begin{array}{l}
u_0=\frac{\alpha b h^3 \theta_1 x (2 E_0+E_1 h)+12 P \left(2 x-L_c e^{-\frac{L+x}{L_c}} \left(e^{\frac{x}{L_c}}-1\right) \left(e^{1/c }+e^{\frac{x}{L_c}}\right)\right)}{12 b h (2 E_0+E_1 h)}\\
w=\frac{(2 E_0+E_1 h) e^{-\frac{L+x}{L_c}} \left(e^{\frac{x}{L_c}} \left(e^{\frac{1}{c}} \left(180 M \left(L_c^2-L_c x+x^2\right)-\alpha b E_1 h^5 \theta_1 x^2\right)+180 L_c M (L_c+x)\right)-180 L_c^2 M ( e^{\frac{1}{c}} - e^{\frac{2 x}{L_c}})\right)}{10 b h^3 \left(6 E_0^2+6 E_0 E_1 h+E_1^2 h^2\right)}.
\end{array}
\end{equation}
Note that in the absence of the external loads, $P=0$ and $M=0$, the nonlocal character of solutions disappears.

\begin{figure}
 	\centering
 	\includegraphics[scale=0.45]{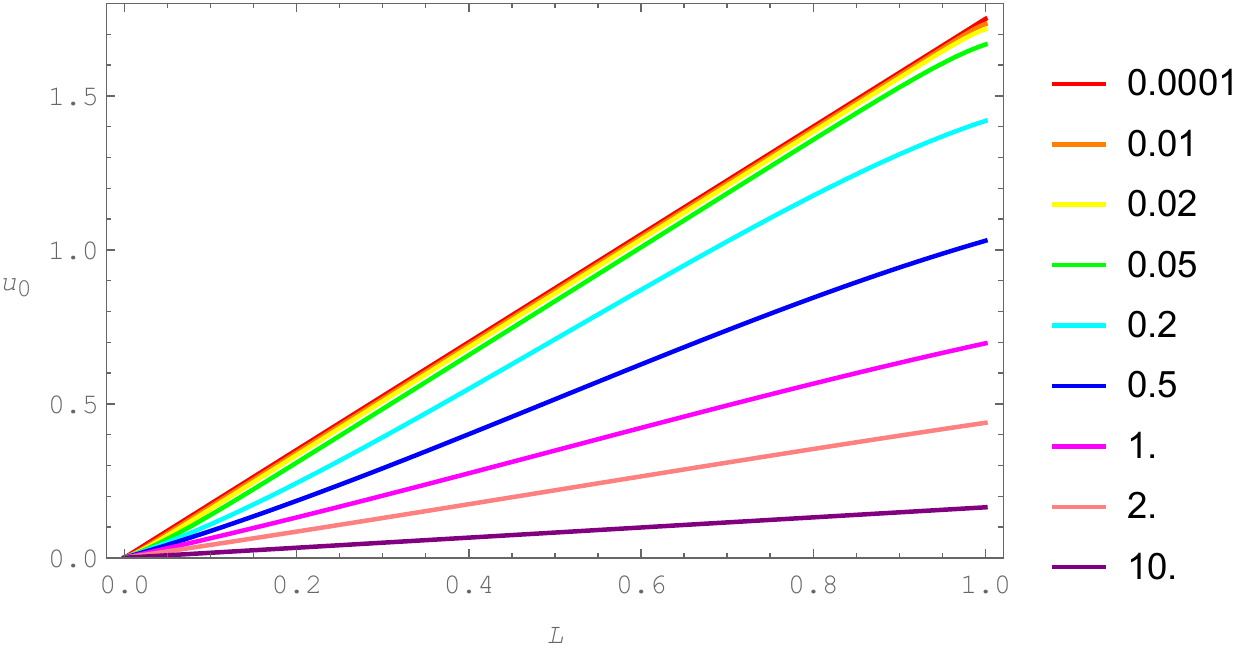}
 	\caption{Gradual stiffening of the axial displacements for $c \in \left\lbrace 0.0001, 0.01, 0.02, 0.05, 0.2, 0.5, 1.0, 2.0, 10\right\rbrace$, cantilever beam.}
 	\label{fig:Ex3_2_u0}
 \end{figure}

\begin{figure}
	\centering
	\includegraphics[scale=0.45]{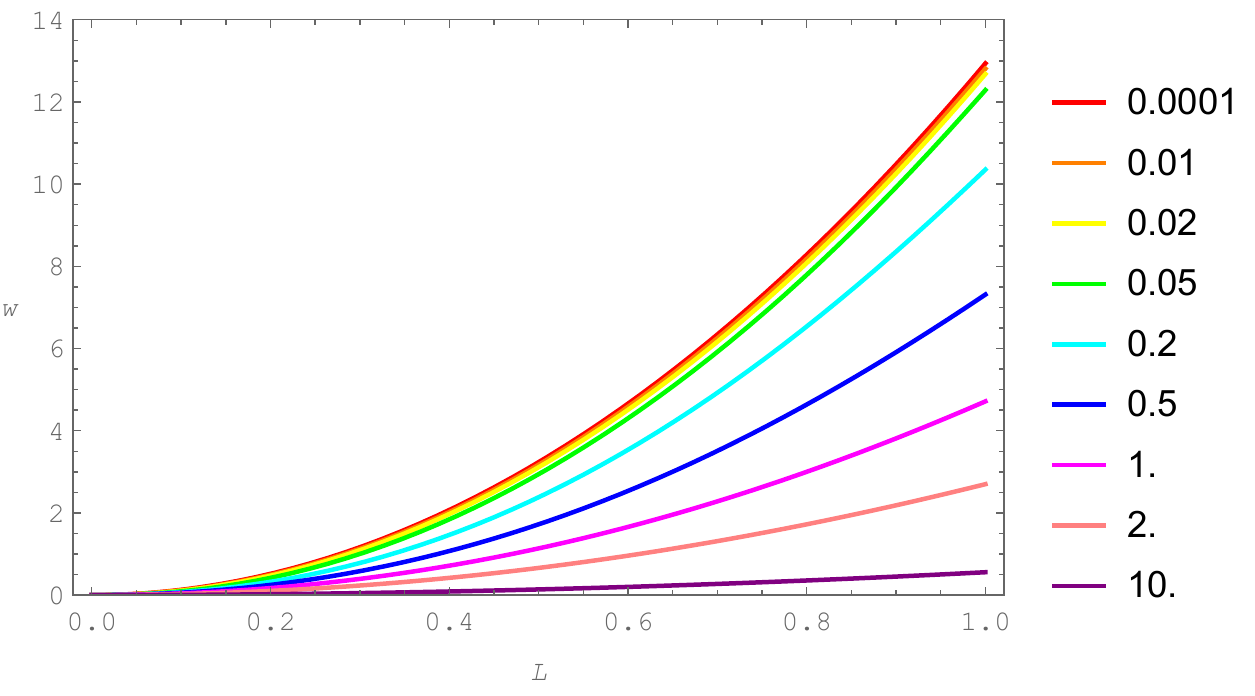}
	\caption{Gradual stiffening of the transverse displacements for $c \in \left\lbrace 0.0001, 0.01, 0.02, 0.05, 0.2, 0.5, 1.0, 2.0, 10\right\rbrace$, cantilever beam.} 
	\label{fig:Ex3_3_w}
\end{figure}

To graphically illustrate solutions, the following values are selected: $L = 1, \alpha = 1, \theta_1 = 0.1, b = 1, h = 1, P = 1, M = 1$. Both solutions were evaluated for the small size parameters $c \in \left\lbrace 0.0001, 0.01, 0.02, 0.05, 0.2, 0.5, 1.0, 2.0, 10\right\rbrace$. The results are given in Figs.~\ref{fig:Ex3_2_u0}, \ref{fig:Ex3_3_w}. Both displacements show the stiffening effect with the increase in the small size parameter.

For the purpose of comparison, the solution for the transverse displacement for the case when the shift of the neutral surface is neglected, $z_0=0$ is also reported:
\begin{equation}\label{eq:Ex3_6}
\begin{array}{l}
w_{z_0}=\frac{3 e^{-\frac{L+x}{L_c}} \left(e^{\frac{x}{L_c}} \left(e^{\frac{1}{c}} \left(80 M \left(L_c^2-L_c x+x^2\right)-\alpha b E_1 h^5 \theta_1 x^2\right)+80 L_c M (L_c+x)\right)-80  L_c^2 M (e^{\frac{1}{c}} - e^{\frac{2 x}{L_c}})\right)}{20 b h^3 (2 E_0+E_1 h)}
\end{array}
\end{equation}
so the transverse displacement assumes simpler form if $z_0=0$. The axial displacement is not affected by the neutral surface shift.

\begin{figure}
	\centering
	\includegraphics[scale=0.45]{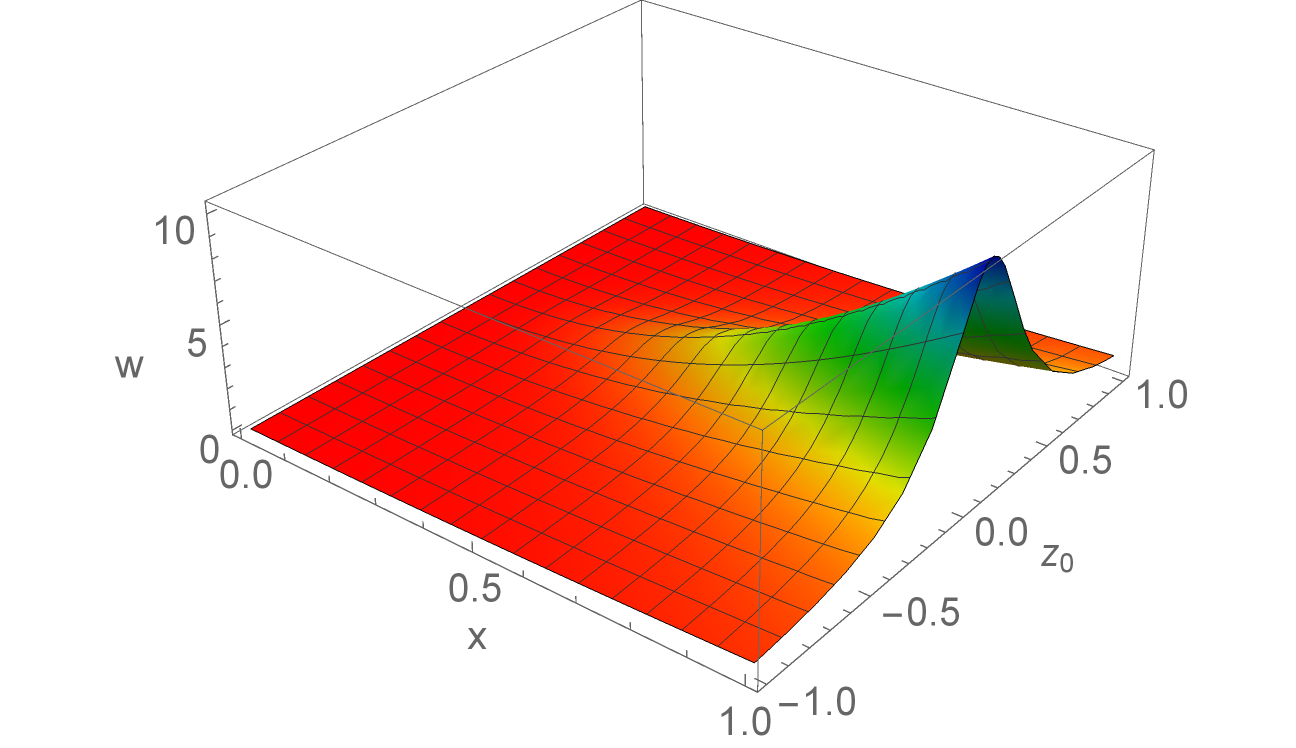}
	\caption{Influence of $z_0$ on the vertical displacement, $c$=0.2.} 
	\label{fig:Ex3_4_z0_1}
\end{figure}

The significance of the neutral surface shift is illustrated in Fig.~\ref{fig:Ex3_4_z0_1}. Increase in beam's stiffness with the increase in $\left| z_0 \right| $ is clearly noticeable. 
 

\subsection{Thermomechanically loaded simply supported FG beam}
A simply supported FG beam is loaded by the distributed axial loading $q_x$ and distributed transverse loading $q_z$. Additionally, beam is exposed to the bilinear temperature field $\Delta \theta=\theta_0 x z$. The beam material is linearly functionally graded like in the previous example $E(z)=E_0+E_1(z+h/2)$, with the same geometry. The coefficient of heat expansion $\alpha$ is taken to be a constant. 

The problem is defined by the differential equations Eqs.~(\ref{eq:U_u0-NLF_Ez1}, \ref{eq:U_w_NLF_15})
\begin{equation}\label{eq:Ex4_1}
\begin{array}{l}
-L_\mathrm{c}^{2} u_0^{(4)} k^\mathrm{AE}  +u_0^{(2)} k^\mathrm{AE} -\Phi^\mathrm{N}_{1} + q_x  =0, \\
-L_\mathrm{c}^{2} w ^{(6)} k^\mathrm{IE} +w ^{(4)} k^\mathrm{IE} +\Phi^\mathrm{M}_{1}- q_z  =0. \\
\end{array}
\end{equation}

The set of boundary conditions follows from Eqs.~(\ref{eq:U_u0-NLF_6},\ref{U_u0-NLF_6a},\ref{eq:U_u0-313},\ref{eq:U_w_NLF_Ez5},\ref{eq:U_w_NLF_Ez6}):
\begin{equation}\label{eq:Ex4_2}
\begin{array}{l}
u_0(0)=0, w(0)=0, w{(L)}=0, \\
\left. -L_\mathrm{c}^{2} u_0^{(3)}k^\mathrm{AE} + u_0^{(1)} k^\mathrm{AE}  -\Phi^\mathrm{N}_{0}  =0 \right| _{x=L},  \\
\left. -L_\mathrm{c}^{2} w ^{(4)} k^\mathrm{IE} +w ^{(2)} k^\mathrm{IE}+\Phi^\mathrm{M}_{0} =0 \right| _{x=0}, \\
\left. -L_\mathrm{c}^{2} w ^{(4)} k^\mathrm{IE} +w ^{(2)} k^\mathrm{IE}+\Phi^\mathrm{M}_{0} =0 \right| _{x=L}, \\
L_\mathrm{c} \left( u_0^{(2)} \int_{\Omega} E \mathrm{d} A-\int_{\Omega} E(\alpha\Delta\theta)^{(1)}\mathrm{d} A\right)  - (u_0^{(1)}\int_{\Omega} E\mathrm{d} A- \int_{\Omega}E\alpha\Delta\theta)\mathrm{d} A  =\left. 0 \right|_{x=0}, \\
L_\mathrm{c} \left( u_0^{(2)} \int_{\Omega} E \mathrm{d} A -\int_{\Omega} E(\alpha\Delta\theta)^{(1)}\mathrm{d} A\right)  + (u_0^{(1)}\int_{\Omega} E\mathrm{d} A- \int_{\Omega}E\alpha\Delta\theta)\mathrm{d} A  =\left. 0 \right|_{x=L}. \\
L_\mathrm{c} \left( - w^{(3)} \int_{\Omega}E(z-z_0)^2 \mathrm{d} A-\int_{\Omega} E (\alpha\Delta\theta)^{(1)}(z-z_0)\mathrm{d} A\right)  \\
- (-w^{(2)} \int_{\Omega} E (z-z_0)^2\mathrm{d} A- \int_{\Omega}E\alpha\Delta\theta(z-z_0)\mathrm{d} A ) =\left. 0 \right|_{x=0}, \\
L_\mathrm{c} \left( - w^{(3)} \int_{\Omega}E(z-z_0)^2 \mathrm{d} A-\int_{\Omega} E (\alpha\Delta\theta)^{(1)}(z-z_0)\mathrm{d} A\right)  \\
+ (-w^{(2)} \int_{\Omega} E (z-z_0)^2\mathrm{d} A- \int_{\Omega}E\alpha\Delta\theta(z-z_0)\mathrm{d} A )  =\left. 0 \right|_{x=L}. \\
\end{array}
\end{equation}
Solving the system of ordinary differential equations just described provides the axial and transverse displacement:
\begin{equation}\label{eq:Ex4_3}
\begin{array}{l}
u_0=\frac{e^{-\frac{L+x}{L_c}} \left(e^{\frac{x}{L_c}} \left(e^{\frac{1}{c}} \left(x^2 \left(\alpha b E_1 h^3 \theta_0-12\right)-12 c  (c +1) L^2+24 L x\right)-12 L_c^2\right)+12 e^{\frac{1}{c}} c  (c +1) L^2+12 L_c^2 e^{\frac{2 x}{L_c}}\right)}{12 b h (2 E_0+E_1 h)},\\
w=\frac{e^{-\frac{L+x}{L_c}}} {6 b h \left(2 E_0 \left(h^2+12 z_0^2\right)+E_1 h \left(h^2-4 h z_0+12 z_0^2\right)\right)} \cdot \\
 \left(e^{\frac{L+x}{L_c}} \left(L^2 x \left(\alpha b h^3 \theta_0 (2 E_0+E_1 (h-2 z_0))+72 c ^2 x\right)\right.\right.\\
\left. \left.+x^3 \left(6 x-\alpha b h^3 \theta_0 (2 E_0+E_1 (h-2 z_0))\right)  +36 c ^3 (2 c +1) L^4+\left(6-72 c ^2\right) L^3 x-12 L x^3\right)\right.\\
 \left.-36 c ^3 (2 c +1) L^4 (e^{\frac{1}{c}}+ e^{\frac{x}{L_c}}-e^{\frac{2 x}{L_c}})\right).\\
\end{array}
\end{equation}

\begin{figure}
	\centering
	\includegraphics[scale=0.45]{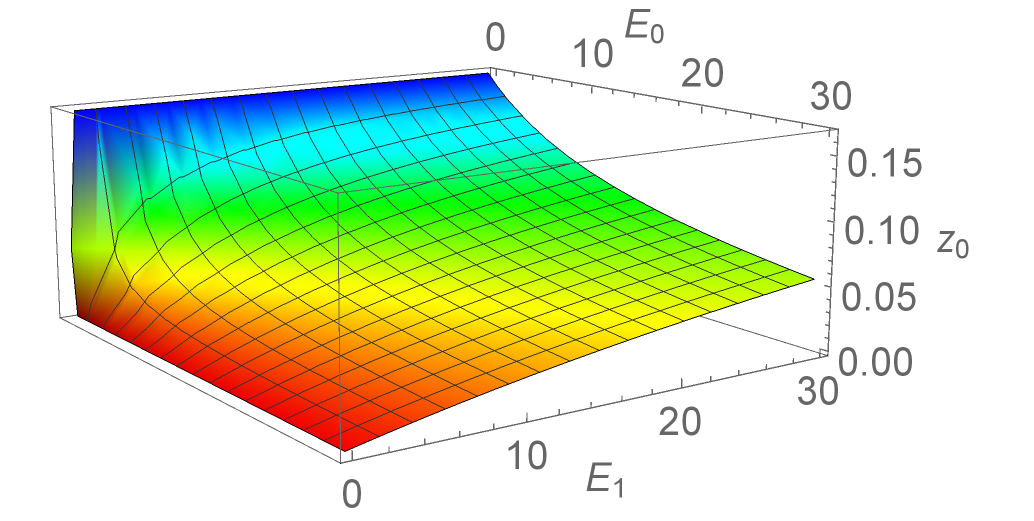}
	\caption{Influence of $E_0$ and $E_1$ on $z_0$} 
	\label{fig:Ex4_2_z0_E}
\end{figure}

Diagrams representing the deformation process are prepared for values: $L = 1, \alpha = 1, \theta_0 = 0.1, b = 1, h = 1, E_0=0.1, E_1=1$. Since the same cross section and material like in Example \ref{sec:Example3} are used, the shift of the neutral surface Eq.~(\ref{eq:Ex3_1}) remains the same: $z_0= \frac{2E_1 I_y}{2E_0 A- b h^2 E_1}$. Dependence of the neutral surface shift of Euler-Young's coefficients $E_0, E_1$ is depicted in Fig.~\ref{fig:Ex4_2_z0_E}. Obviously, for $E_1=0$ the shift is $z_0=0$, while for larger values of $E_0, E_1$ converges toward $z_0 \approx 0.6$.

\begin{figure}
	\centering
	\includegraphics[scale=0.45]{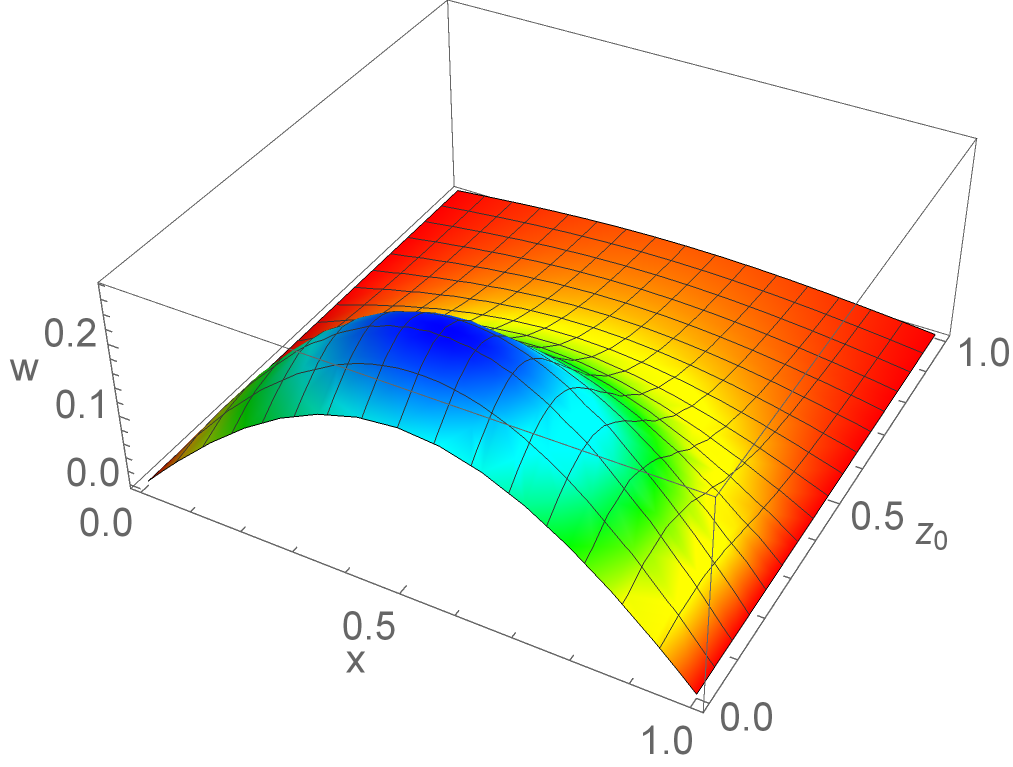}
	\caption{Influence of $z_0$ on the transverse displacement, $c$=0.2.} 
	\label{fig:Ex4_1_z0}
\end{figure}

Obtained displacement distributions Eq.~(\ref{eq:Ex4_3})$_1$ show that the axial displacement is independent of the neutral shift. This is not the case with the transverse displacement, Fig.~\ref{fig:Ex4_1_z0}. As the shift of the neutral surface increases, the beam becomes more stiffer, asymptotically converging to the rigid beam.

\begin{figure}
	\centering
	\includegraphics[scale=0.45]{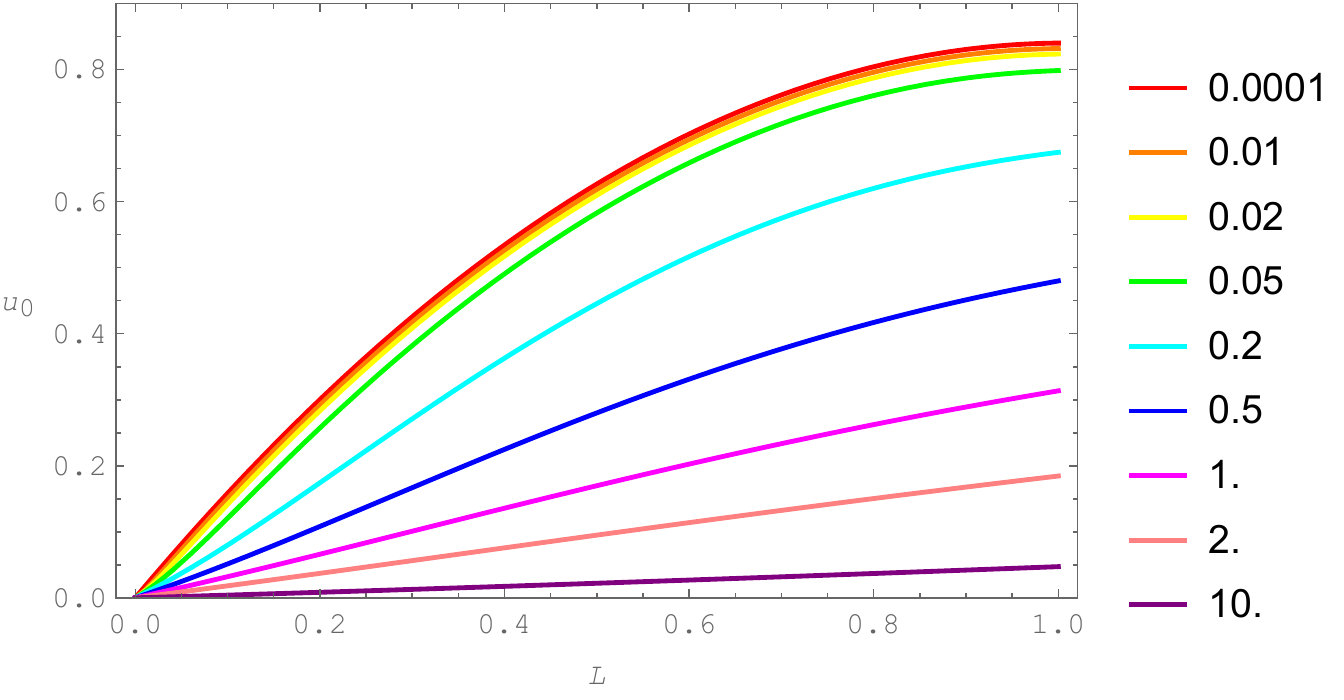}
	\caption{Gradual stiffening of the axial displacements for  $c \in \left\lbrace 0.0001, 0.01, 0.02, 0.05, 0.2, 0.5, 1.0, 2.0, 10\right\rbrace$, simply supported beam.}
	\label{fig:Ex4_3_u0}
\end{figure}

\begin{figure}
	\centering
	\includegraphics[scale=0.45]{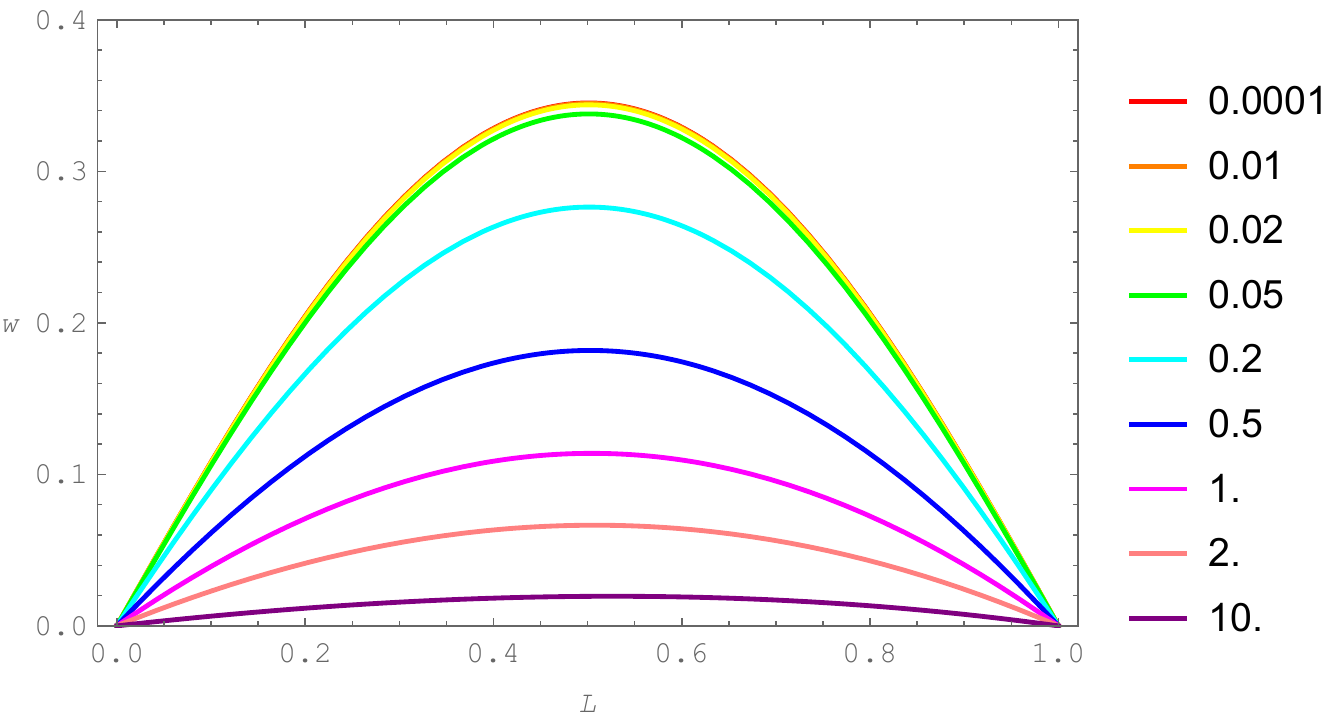}
	\caption{Gradual stiffening of the transverse displacements for $c \in \left\lbrace 0.0001, 0.01, 0.02, 0.05, 0.2, 0.5, 1.0, 2.0, 10\right\rbrace$, simply supported beam.} 
	\label{fig:Ex4_4_w}
\end{figure}

Like before, solutions were evaluated for the same set of small size parameters $c \in \left\lbrace 0.0001, 0.01, 0.02, 0.05, 0.2, 0.5, 1.0, 2.0, 10\right\rbrace$. The results are given in Figs.~\ref{fig:Ex4_3_u0}, \ref{fig:Ex4_4_w} repeating behaviour observed in Example \ref{sec:Example3} - displacements demonstrate the stiffening effect as $c$ is increased. Hence, such a beam shows the standard nonlocal behaviour.

\section{Conclusions}
\label{Conclusions}
The paper at hand extended the existing nonlocal Bernoulli-Euler homogeneous beam formulation to functionally graded materials. The presented formulation is based on the consistent thermodynamic formulation emanating from a suitably selected potential. Most significant findings and results are summarized below.
\begin{itemize}
	\item A thermodynamically consistent formulation is employed to develop the proposed formulation, based on a proper Gibbs potential, the stress-driven integral model for functionally graded nanobeams. Minimization of the total energy potential provides underlying differential equations and boundary conditions governing the relevant thermoelastostatic problem.  
	\item The shift of neutral surface is included into the formulation. It is demonstrated that this effect can have a profound influence on deformation, especially in the presence of small-size phenomena, and should not be disregarded.
	\item It is noticed that in the case of doubly-clamped nonisothermal beams, for the lower values of the nonlocal parameters, a softening behaviour compared to the local case is exhibited. However, as the nonlocal parameter is further increased, the beam starts to stiffen, asymptotically converging to the local solution. The similar conclusion is reached in \cite{Canadija2018} in the special framework of elastically homogeneous materials. 
\end{itemize}

\section{Acknowledgements} \label{Acknowledgements}
This work has been fully supported by the University of Rijeka under the project number uniri-tehnic-18-37 - Mechanical behaviour of nanostructures. This support is gratefully acknowledged

\section{Data availability}
Apart from the analytical results presented in the manuscript, there are none other data appropriate for sharing.

\bigskip

\noindent


\bibliographystyle{elsarticle-num}

\bibliography{SmallSizeParameter}

\end{document}